\documentclass[%
 reprint,
 amsmath,amssymb,
prl,
floatfix,
]{revtex4}
\usepackage{dcolumn,xcolor}

\usepackage{graphicx}
\usepackage{dcolumn}
\usepackage{bm}

\begin{document}


\title{Optical Control of Skyrmion Trajectories via Skyrmion Number Currents}

\author{Emir Syahreza Fadhilla$^{1}$}
\email{emir002@brin.go.id}
\author{M Shoufie Ukhtary$^1$}
\email{msho001@brin.go.id}
\author{Ardian Nata Atmaja$^1$}%
\email{ardi002@brin.go.id}
\author{Bobby Eka Gunara$^{2}$}
\email{bobby@itb.ac.id}

\affiliation{%
 $^1$Research Center for Quantum Physics, National Research and Innovation Agency (BRIN)\\
 Kompleks PUSPIPTEK Serpong, Tangerang 15310, Indonesia,
}
\affiliation{%
 $^2$Theoretical High Energy Physics Research Division,\\
 Faculty of Mathematics and Natural Sciences, Institut Teknologi Bandung,\\
 Jl. Ganesha no. 10 Bandung, Indonesia, 40132,
}%


\date{\today}

\begin{abstract}
We propose a mechanism to control magnetic Skyrmion motion by generating a Skyrmion number current. This current is produced and controlled via explicitly time-dependent Hamiltonian containing a Zeeman term arising from the interaction between the spin system and circularly polarized light. We find that Skyrmion dynamics exhibit a limit cycle in momentum space, whose properties are determined only by physical parameters: the magnitude of external magnetic field, the Heisenberg coupling, and the Gilbert damping constant. This method provides a theoretical explanation for the topological origin of optically controlled Skyrmion motion, and it opens the possibility for efficient, low-dissipation Skyrmion control using Skyrmion number currents as an alternative to electric currents. 
\end{abstract}

\maketitle

\section{Introduction}The study of the equations of motion (EOM) of local defects in magnetic systems is a central topic in spintronics, as it underpins strategies for controlling magnetic states~\cite{zhang2020skyrmion}. Among these defects, a particularly important class is the topological defects, whose configurations are protected from the vacuum by topological stabilisation, which makes them more robust against perturbations. This protection originates from the conservation of a topological charge associated with the nontrivial topology of the spin texture in real space~\cite{Manton:2004tk,luo2018reconfigurable,je2020direct,xu2023reconfigurable}. It is known that the EOM of such defects are influenced by both the surrounding electric current and the external magnetic field profile, which can be exploited to control their motion~\cite{tatara2008microscopic,zang2011dynamics,casiraghi2019individual}. However, scattering between the electric current and the lattice leads to significant energy dissipation, reducing the efficiency of electric current-based control methods~\cite{fujita2017encoding,fujita2017ultrafast,guan2023optically}.

One of the physically interesting topological defects in two- and three-dimensional systems is the Skyrmion, a quasiparticle which was first proposed in models of strongly interacting particles \cite{Skyrme:1961vq,Skyrme:1962vh,Piette:1994mh,Piette:1994ug}. The Skyrmion solutions are then found in the models of two-dimensional magnetic systems, which were considered as metastable states \cite{Polyakov:1975yp}, but later it was found that such solutions can be stabilized by the competition between magnetostatic energy and Dzyaloshinskii–Moriya (anti-symmetric) exchange interaction (DMI) \cite{1989JETPBogdanov}.

A Skyrmion possesses a topological invariant, \(Q\equiv \int\textbf{n}\cdot\left(\partial_x\textbf{n}\times\partial_y\textbf{n}\right)~d^2x/(4\pi)\in\mathbb{Z}\), which is known as the Skyrmion number \cite{nagaosa2013topological,han2017skyrmions}, where \(\textbf{n}(t,\textbf{r})\equiv\textbf{n}\) is the spin orientation at time \(t\) and position \(\textbf{r}\). The Skyrmion number is known to be a conserved charge during the time evolution of Skyrmion unless the Skyrmion breaks down through certain processes \cite{zhang2020skyrmion,xu2023reconfigurable}.
Using the definition of Skyrmion number, we can define the Skyrmion number density, \(q\), and current density, \(\textbf{j}\),
\begin{equation}\label{DefSkyrNumDensities}
    q\equiv\frac{1}{4\pi}\textbf{n}\cdot\left(\partial_x\textbf{n}\times\partial_y\textbf{n}\right),~\textbf{j}\equiv\frac{1}{4\pi}\begin{bmatrix}
        -\textbf{n}\cdot\left(\partial_t\textbf{n}\times\partial_y\textbf{n}\right)\\\textbf{n}\cdot\left(\partial_t\textbf{n}\times\partial_x\textbf{n}\right)
    \end{bmatrix}.
\end{equation}
The Skyrmion number current density defined above is chosen such that we have a continuity equation for Skyrmion number
\begin{equation}\label{QConserv}
    \partial_tq+\nabla\cdot\textbf{j}=0\Leftrightarrow \frac{dQ}{dt}=0.
\end{equation}
This implies that the Skyrmion number is preserved, by definition \eqref{DefSkyrNumDensities}. It is worth noting that the continuity equation \eqref{QConserv} is equivalent to the Faraday's law of induction of the emerging electromagnetic fields in magnetic systems \cite{barnes2007generalization}, namely \(\partial_t \textbf{b}+\nabla\times\textbf{e}=0\),
where the emergent magnetic field is \(\textbf{b}=\textbf{n}\cdot\left(\partial_x\textbf{n}\times\partial_y\textbf{n}\right)\hat{z},\) and emergent electric field is \(\textbf{e}=\textbf{n}\cdot\left(\partial_x\textbf{n}\times\partial_t\textbf{n}\right)\hat{x}+\textbf{n}\cdot\left(\partial_y\textbf{n}\times\partial_t\textbf{n}\right)\hat{y}\), directly derived from the time and spatial dependence of \(\textbf{n}\). This emergent Faraday's law has been experimentally confirmed \cite{lee2009unusual,neubauer2009topological,schulz2012emergent} which implies that the Skyrmion number is indeed a conserved charge.

The influence of the Skyrmion number current on the equations of motion of magnetic Skyrmions has been investigated theoretically, and it is known to compete with the effect of the electric current~\cite{papanicolaou1991dynamics,stone1996magnus}. In the absence of an electric current, the Skyrmion's velocity is therefore strongly governed by the Skyrmion number current. Since the Skyrmion number depends on $\partial_t\mathbf{n}$, this current can be controlled by adjusting the dynamics of $\mathbf{n}$, for example, through the application of an external magnetic field~\cite{guan2023optically,kazemi2024all,titze2024all}. We are going to show that this approach enables indirect control of the Skyrmion trajectory via time-dependent external fields acting through the Skyrmion number current. Compared to electric-current-based methods, Skyrmion-number-current-based control offers a wider range of possible mechanisms, with the potential for higher energy efficiency.

We propose that a significant effect of topology on the Skyrmion motion can be observed in the following setup for an isolated Skyrmion in a medium that is free from electric current. Let \(\textbf{n}=(\sin\Theta\cos\Phi,\sin\Theta\sin\Phi,\cos\Theta)\), be the orientation of the spin at position \(\textbf{r}\), with \(\Theta\in[0,\pi]\) and \(\Phi\in[0,2\pi)\). In general, formulating the defect's EOM is done by introducing collective coordinate motion, \(\textbf{R}(t)\), as a time- dependent spatial translation, \(\textbf{n}(t,\textbf{r})\rightarrow\textbf{n}(t,\textbf{r}-\textbf{R}(t))\), such that the time-derivative of \(\textbf{n}\) transforms as \(\partial_t\textbf{n}\rightarrow\partial_t\textbf{n}-(\partial_t\textbf{R}\cdot\nabla)\textbf{n}\). The vector \(\textbf{R}(t)\) is the center of mass of the defect and, for the case of Skyrmion, the center of mass is also the center of Skyrmion where \(\Theta=0\). The Landau-Lifshitz-Gilbert (LLG) equation without spin-transfer torque, that is given by
\begin{equation}
    \partial_t\textbf{n}=\gamma\left(\textbf{n}\times\frac{\delta H}{\delta\textbf{n}}+\alpha\textbf{n}\times\partial_t\textbf{n}\right),
\end{equation}
where \(\gamma\) is the gyromagnetic ratio and \(\alpha\) is the Gilbert damping constant, also transforms under collective coordinate transformation
\begin{eqnarray}\label{SpatDependent}
    &&\textbf{n}\cdot\left(\partial_t\textbf{n}\times\partial_i\textbf{n}\right)-\textbf{n}\cdot\left((\partial_t\textbf{R}\cdot\nabla)\textbf{n}\times\partial_i\textbf{n}\right)\\&=&\gamma \partial_i\textbf{n}\cdot\frac{\delta H}{\delta \textbf{n}}+\gamma\alpha\partial_i \textbf{n}\cdot\partial_t\textbf{n}-\gamma\alpha(\partial_t\textbf{R}\cdot\nabla)\textbf{n}\cdot\partial_i\textbf{n},\nonumber
\end{eqnarray}
with index \(i\in \{x,y\}\).
Here, \(H\) is the Hamiltonian density of the system, which contains the magnetic energies from the interaction between spins and external fields.
In the equation for \(\textbf{R}(t)\) above, we find two topological quantities, namely the Skyrmion number density, in the second term in LHS, and the components of the Skyrmion number current density, that is, the first term on the LHS. We can further remove the spatial dependence of Eq. \eqref{SpatDependent} by introducing the strain tensor \cite{manton1987,Fadhilla:2021jiz}, \(\mathcal{D}\), whose components are \(\mathcal{D}_{ij}\equiv\partial_i\textbf{n}\cdot\partial_j\textbf{n}\),
which allows us to introduce the following three quantities that are proportional to the integrals of Skyrmion number current and strain tensor,
\begin{equation}\label{DynamicalParameters}
\textbf{J}_Q=\int\textbf{j}~d^2x,~\frac{\mathcal{G}}{\gamma\alpha}=\int\mathcal{D}~d^2x,~\frac{\textbf{D}}{\gamma\alpha}\equiv\int\begin{bmatrix}
    \partial_t\textbf{n}\cdot\partial_x\textbf{n}\\
    \partial_t\textbf{n}\cdot\partial_y\textbf{n}
\end{bmatrix}d^2x,
\end{equation}
such that we can rewrite the EOM into the following form,
\begin{equation}\label{SpatIndependent}
\nabla_\textbf{R}V(\textbf{R})+\mathcal{G}\cdot\partial_t\textbf{R}-4\pi \hat{z}\times\left(Q\partial_t\textbf{R}+\textbf{J}_Q\right)=\textbf{D}.
\end{equation}
The tensor \(\mathcal{G}\) is well-known as the Skyrmion's shape factor \cite{han2017skyrmions}, and the vector \(\textbf{D}\) is the driving term which arise from the time dependence of \(\textbf{n}\). The effect of the Skyrmion number and the shape factor on the Skyrmion's motion is already well-understood since it defines the Hall angle of the Skyrmion \cite{stone1996magnus,jiang2017direct,chen2017skyrmion,litzius2017skyrmion}. However, the effect of Skyrmion number current is often neglected due to the assumption that the Skyrmion profile is in steady state, i.e. \(\partial_t\textbf{n}\approx0\) at large \(t\). When \(\partial_t\textbf{n}\approx0\) is assumed, both \(\textbf{J}_Q\) and \(\textbf{D}\) are negligible and Eq. \eqref{SpatIndependent} becomes the standard Thiele's equation for magnetic Skyrmion \cite{thiele1973steady,lobanov2024dynamics}. In the case of a spin system strongly interacting with an external time-dependent perturbation, such a steady-state approximation cannot always be satisfied. 

\section{Time Dependent Skyrmion Profile Under Circularly-Polarized Light} Let us demonstrate a simple mechanism where we can manipulate the Skyrmion number current. Consider a spin system with Heisenberg (symmetric) exchange interaction that is affected by a circularly polarized light, whose magnetic field is given by \(\textbf{B}(t)=B_0(\cos(\omega t),\sin(\omega t))\), through the Zeeman effect. This type of light is chosen such that the external perturbation possesses non-zero angular momentum \cite{fujita2017encoding}, which in this case is in the form of spin angular momentum (SAM). The Hamiltonian density of this system is given by
\begin{equation}
    H=J\partial_i\textbf{n}\cdot\partial_i\textbf{n}-\textbf{B}(t)\cdot\textbf{n},
\end{equation}
where \(J\) is the Heisenberg coupling constant, not to be confused with the definition \eqref{DynamicalParameters}, which is the spatial integral of Skyrmion number current. To calculate the Skyrmion number current generated by this type of Hamiltonian, we need to first find the Skyrmion profile \(\textbf{n}(t,\textbf{r})\). 

When \(\textbf{B}(t)\) is turned off, the solution for this model can be given in closed form and is known as the Belavin-Polyakov (BP) solution \cite{Polyakov:1975yp}. The Skyrmion satisfying BP solution is a meta-stable Skyrmion which can only be stabilized through conservation of Skyrmion number when thermal fluctuation is absent. As such, the derived Skyrmion profile, \(\textbf{n}(t,\textbf{r})\), assumes zero temperature, such that the system is exactly at the lowest energy state. We then apply the perturbation method, considering \(\textbf{B}(t)\) as a perturbation. The Hamiltonian \eqref{Hamiltonian} is chosen such that the unperturbed Skyrmion profile is in closed form. This choice of setup allows for analytic treatment and enables us to isolate the role of the skyrmion number current in driving Skyrmion motion.

However, to apply the perturbation method, \(B_0\) cannot be too large since the Skyrmion must be stable. As such, the energy brought to the system by the light cannot exceed the energy loss to the damping over a certain period. Let \(U\equiv\int H d^2x\) be the total energy of the system, then \(dU/dt\leq0\). This criterion ensures that the system always evolves into a minimum-energy configuration. By assuming that there exists a stable Skyrmion radius, \(\lambda_0\), satisfying the BP solution, we have the following upper bound for \(B_0\), which relates the magnetic field strength with the damping constant and the Heisenberg coupling (See appendix Sect. A1 for details), namely
\begin{equation}\label{BBound}
    \omega B_0 <\frac{\alpha\gamma^2}{1+\alpha^2\gamma^2}\frac{16J^2}{\lambda_0^4}.
\end{equation}
We would like to note that the upper bound \eqref{BBound} can be refined using a more rigorous stability analysis, for example, by introducing noise or thermal fluctuations to the model. However, it will require interaction terms other than the Heisenberg interaction in the Hamiltonian to stabilize the Skyrmion, for example, the DMI \cite{BOGDANOV1994255,roessler2006spontaneous}. Such an approach and its implications are beyond the scope of this work and will be discussed elsewhere.

\begin{figure}[]
    \includegraphics[width=0.4\linewidth]{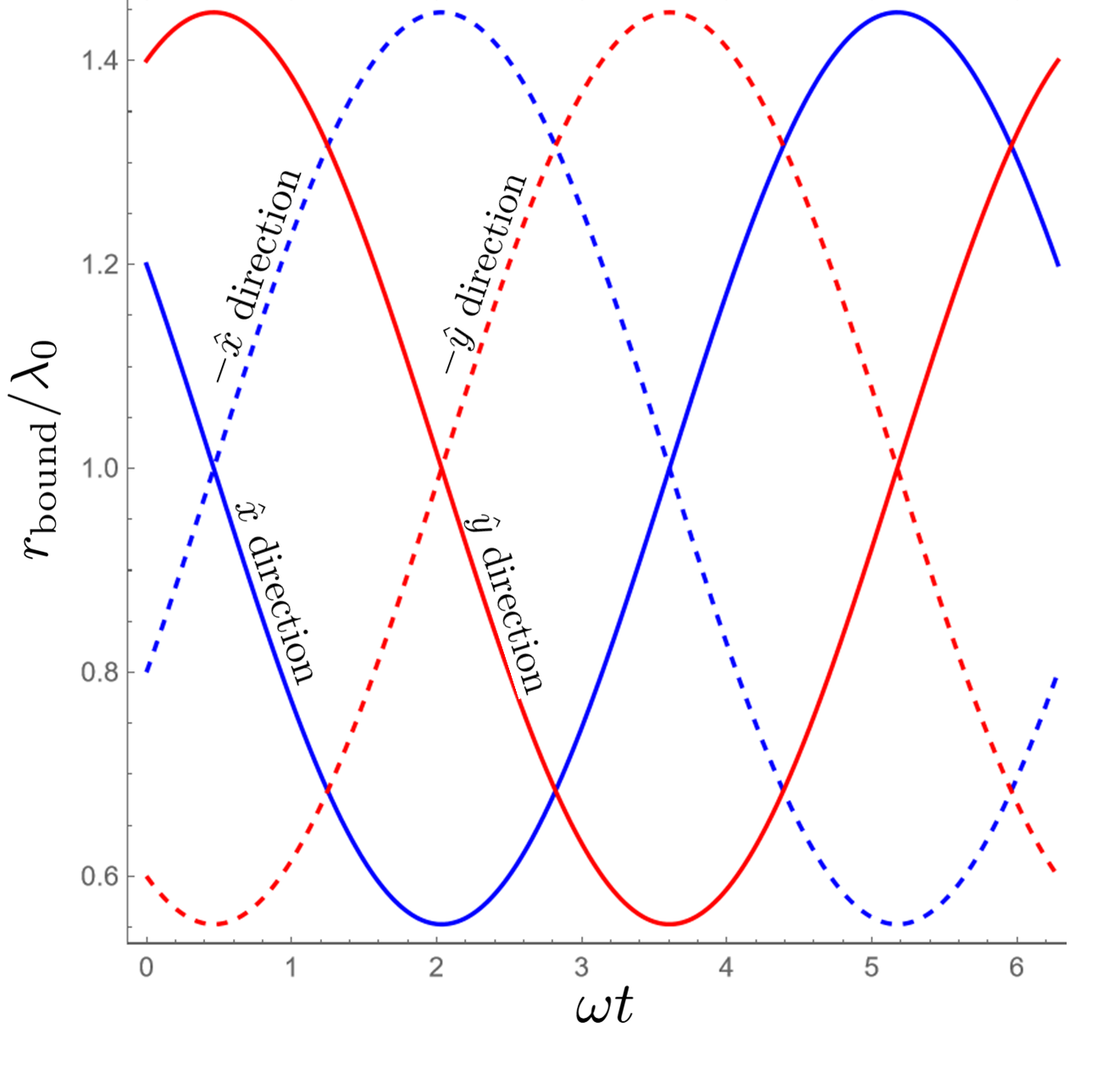}
    \caption{The ratio between the distance from center to boundary, \(r_{\text{bound}}\), and the unperturbed Skyrmion radius, plotted against \(\omega t\) for a single period, from \(t=0\) up to \(t=2\pi/\omega\). This plot shows the anisotropic breathing of the \(Q=1\) Skyrmion with different phases in each direction.}
    \label{fig:RBoundVsTime}
\end{figure}
The perturbed solution can be assumed to be a breathing Skyrmion in the BP form that is explicitly given by,
\begin{equation}\label{BPtypeSol}
    \Theta\equiv2\tan^{-1}\left(\frac{r}{\lambda(t,\varphi)}\right),~~~\Phi\equiv\Phi(t,\varphi),
\end{equation}
where \((r,\varphi)\) are the polar coordinates on the 2D plane.
Under the perturbation method, the functions are expanded in terms of the perturbation parameter \(\epsilon\), as \(\lambda(t,\varphi)=\lambda_0+\sum_{k=1}^\infty\epsilon^k\tilde{\lambda}_k(t,\varphi),\) and \(\Phi(t,\varphi)=Q\varphi+\varphi_0+\sum_{k=1}^\infty\epsilon^k\tilde{\Phi}_k(t,\varphi),\)
while the perturbed LLG equation that is to be solved is given by,
\begin{equation}\label{PerturbedLLG}
    \partial_t\textbf{n}=-\gamma~\textbf{n}\times\left[\epsilon\textbf{B}(t)+2J\nabla^2\textbf{n}\right]+\alpha\gamma~ \textbf{n}\times\partial_t\textbf{n}.
\end{equation}
Let \(\textbf{n}_0\) be the unperturbed solution where \(\epsilon\) is set to zero, then \(\textbf{n}_0\) satisfies \(\textbf{n}_0\times\nabla^2\textbf{n}_0=0\), that is the steady state limit of the unperturbed Eq. \eqref{PerturbedLLG}. We can imagine the solution \eqref{BPtypeSol} as the Skyrmion profile whose boundary, that is the set of points satisfying \(\hat{z}\cdot\textbf{n}(t,\textbf{r})=0\), is stretched in several directions, hence the name breathing Skyrmion. The parameter \(\varphi_0\) defines the helicity of the Skyrmion, where \(\varphi_0=0\) (or \(\pi\)) corresponds to the Neel (or Anti-Neel) Skyrmion, while \(\varphi_0=\pm \pi/2\) corresponds to the Bloch Skyrmion.  The parameter \(\lambda_0\) is the radius of the Skyrmion, which is the distance from the center to the boundary of the unperturbed solution. As such, the perturbation method should be evaluated near \(r=\lambda_0\) where the boundary is stretched.

Up to first order perturbation, the solution of \(\lambda\) and \(\Phi\) are of the following form 
\begin{eqnarray}\label{lambdat}
    \lambda(t,\varphi)&=&\lambda_0+\lambda_0A^+_\lambda\cos\left(Q\varphi+\varphi_0-\omega t\right)\nonumber\\&&+\lambda_0A^-_\lambda\sin\left(Q\varphi+\varphi_0-\omega t\right)~,\\
    \Phi(t,\varphi)&=&Q\varphi+\varphi_0+A^+_\Phi\cos\left(Q\varphi+\varphi_0-\omega t\right)\nonumber\\&&+A^-_\Phi\sin\left(Q\varphi+\varphi_0-\omega t\right)~.\label{phit}
\end{eqnarray}
where the amplitudes \(A^\pm_\rho\equiv A^\pm_\rho(\alpha\gamma,\eta_1,\eta_2)\), \(\rho\in\{\lambda,\Phi\}\), depend on the damping constant and two dimensionless physical parameters of the system, defined as follows (See Appendix Sect. A2 for details) 
\begin{equation}
    \eta_1=\frac{\gamma B_0}{\omega}~,~~~\eta_2=2\frac{\gamma J}{\omega\lambda_0^2}~.
\end{equation}
The ratio \(\eta_1/\eta_2=B_0\lambda_0^2/(2J)\), is the measure of the external magnetic field's strength compared to the magnetic interaction between the spins \cite{kasteleijn1956constant}. The upper bound for \(B_0\) in \eqref{BBound}, can be recast into relation between \(\eta_1\) and \(\eta_2\) as \(\eta_1<4\alpha\gamma\eta_2^2/(1+\alpha^2\gamma^2)\). The solutions \eqref{lambdat} and \eqref{phit} represent a Skyrmion profile with anisotropic breathing with different breathing phases in each direction. As demonstrated in Fig. \ref{fig:RBoundVsTime}, the phase difference between adjacent directions is \(\pi/2\) for \(Q=1\) Skyrmion. In this figure, we plot the ratio between \(r_\text{bound}\), the distance from the center to the Skyrmion boundary in a certain direction, and the unperturbed Skyrmion radius, \(\lambda_0\), against the phase, \(\omega t\). \(r_\text{bound}\)'s are computed in the \(\pm\hat{x}\) and the \(\pm\hat{y}\) directions to show how the Skyrmion boundary stretches at the main axes according to this breathing mode. For \(Q>1\), the direction of anisotropic breathing increases proportionally with \(Q\). 

The total Skyrmion number current, \(\textbf{J}_Q\), of the \(Q=1\) breathing Skyrmion has the following oscillatory form
\begin{equation}
    \textbf{J}_Q=\begin{bmatrix}
        J_{Q}^-\\
        J_{Q}^+
    \end{bmatrix}\cos\left(\varphi_0-\omega t\right)+\begin{bmatrix}
        J_{Q}^+\\
        -J_{Q}^-
    \end{bmatrix}\sin\left(\varphi_0-\omega t\right),
\end{equation}
where its components on the Cartesian basis hold non-zero value only for \(Q=\pm1\). We found that time-dependent \(\textbf{n}\) alone cannot guarantee a nonzero \(\textbf{J}_Q\) and that anisotropy is needed (See Appendix Sect. A3). The constants \(J_{Q}^\pm\) above can be written explicitly in terms of \(A^\pm_\rho\), namely \(J^{\pm}_Q=\pi\lambda_0\omega\left[A_\Phi^\mp\pm A_\lambda^\pm\right]/4\).
This total Skyrmion number current, together with the driving term, determines the late-time trajectory of the Skyrmion. When both \(\textbf{J}_Q\) and \(\textbf{D}\) are zero, the Skyrmion velocity will always converge to zero at \(t\rightarrow\infty\) due to the dissipation by the Gilbert Damping.

\begin{figure}[]
    \includegraphics[width=0.25\linewidth]{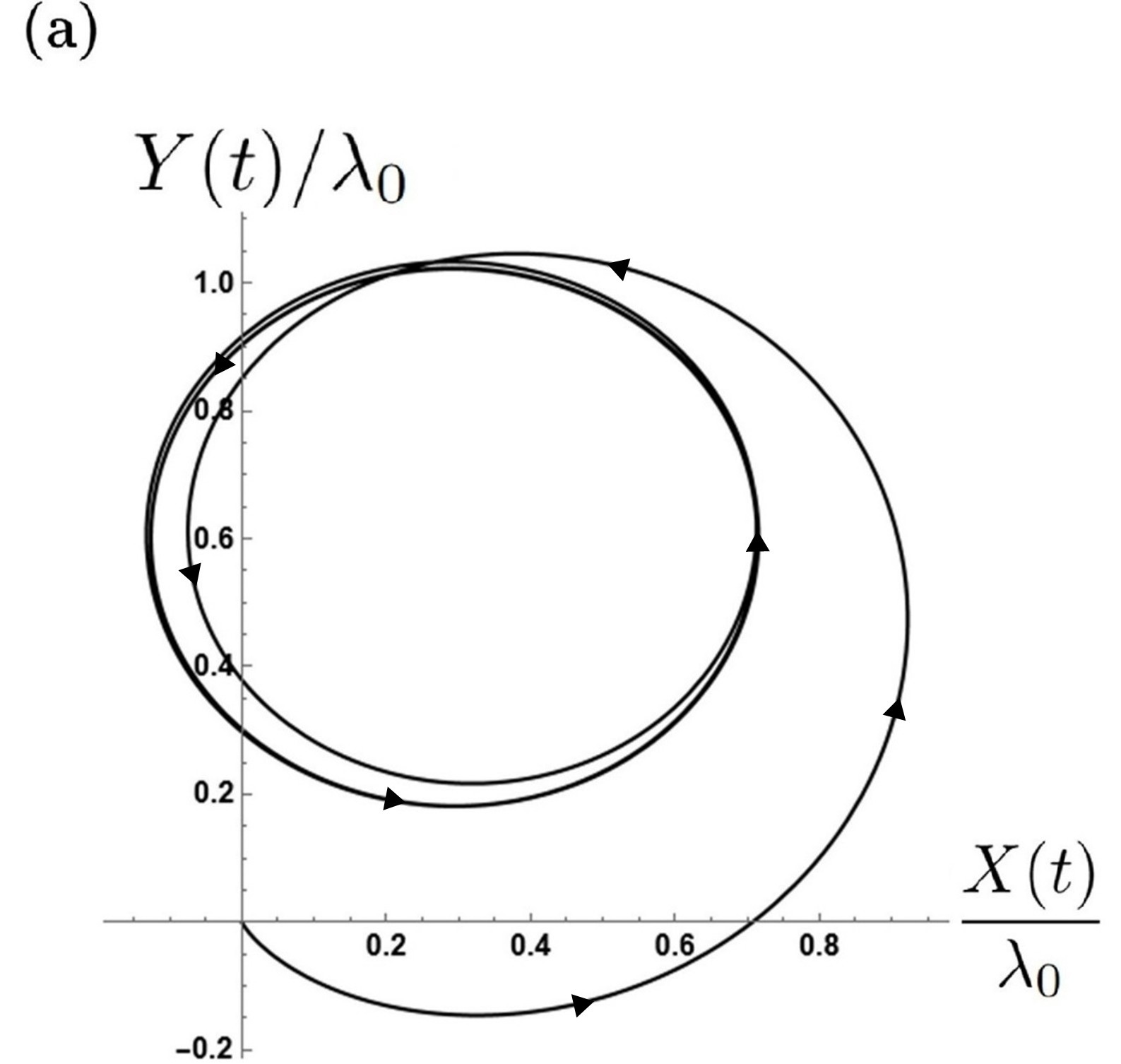}
    \includegraphics[width=0.24\linewidth]{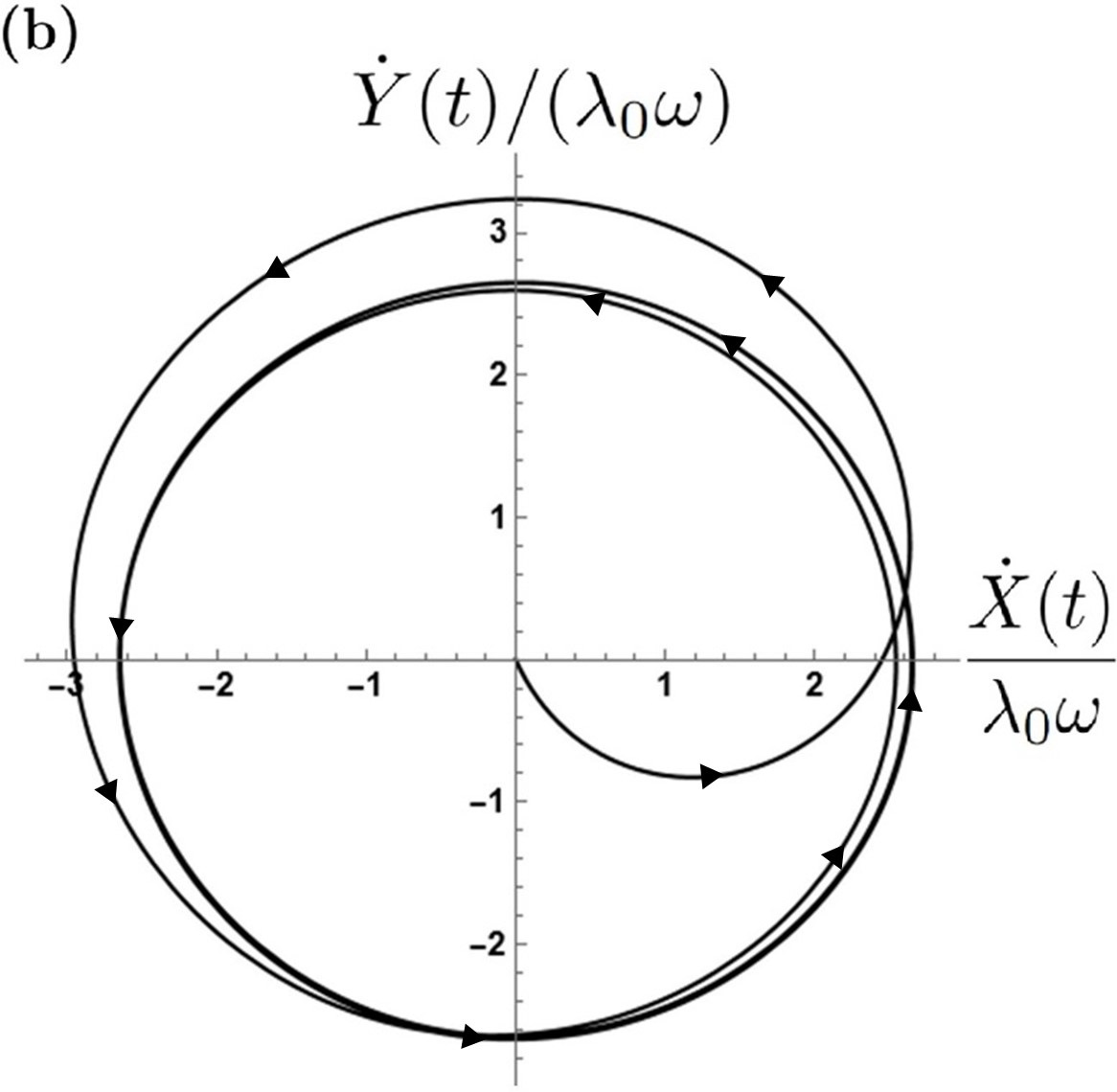}
    \includegraphics[width=0.23\linewidth]{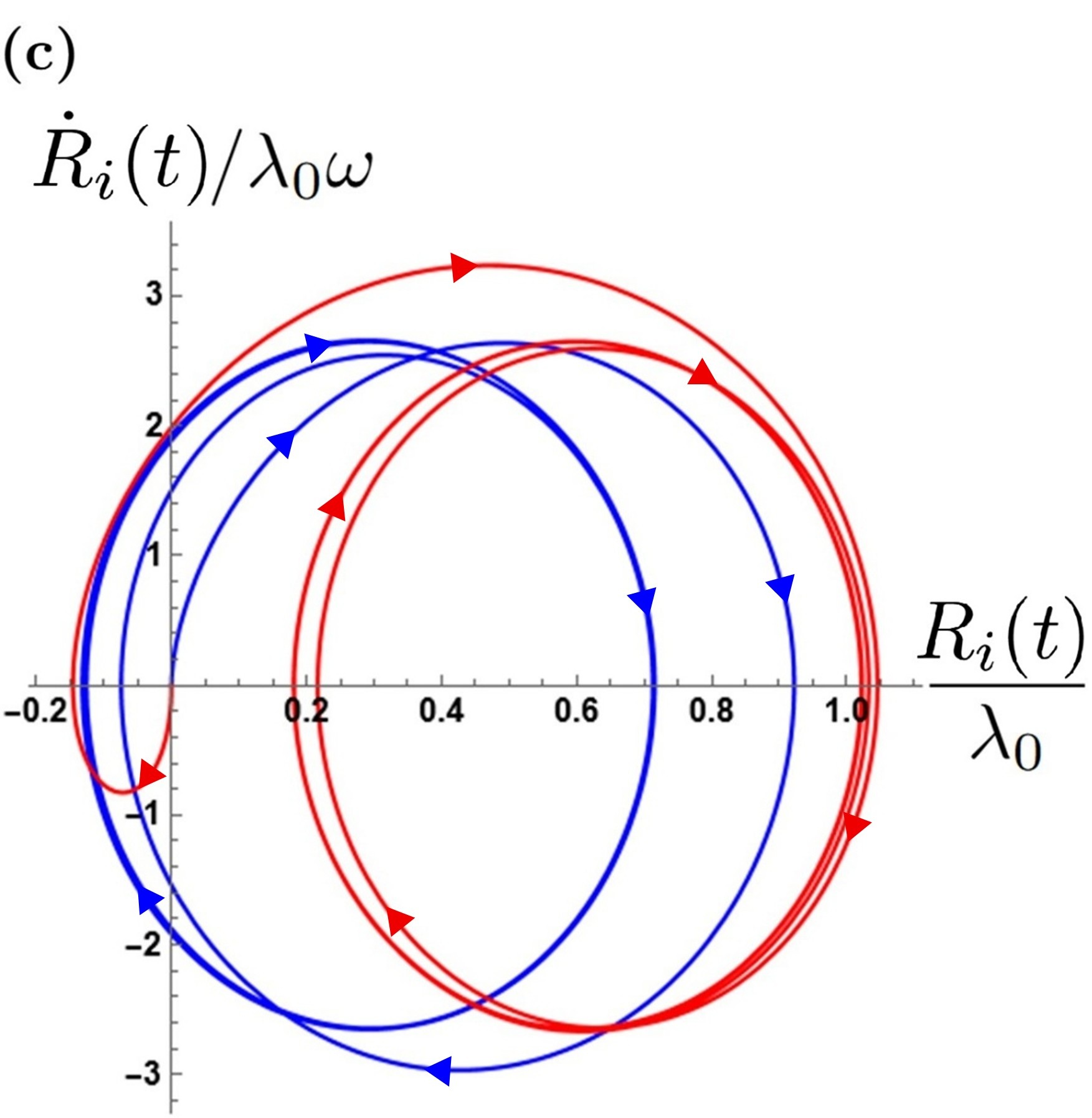}
     \includegraphics[width=0.23\linewidth]{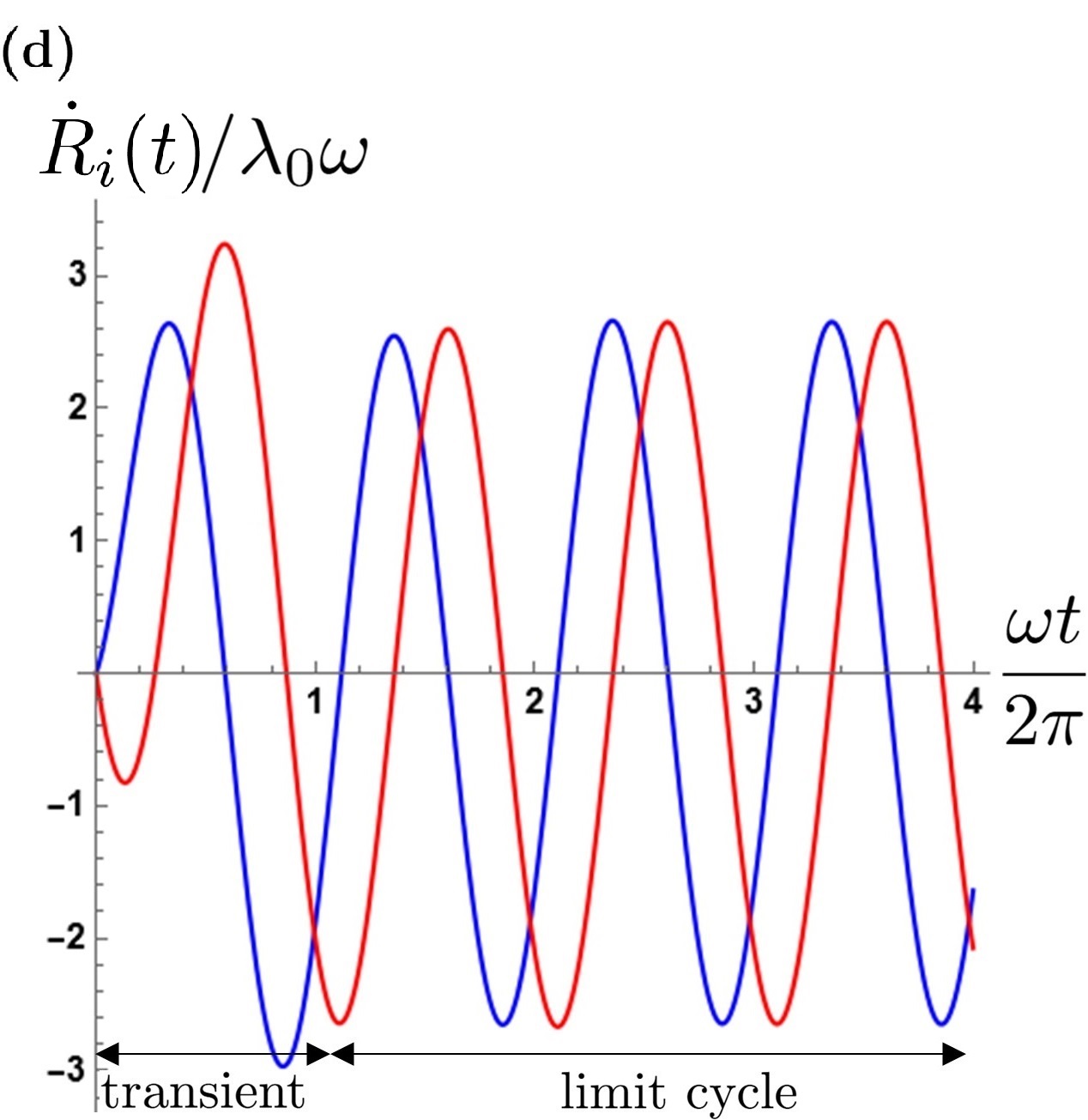}
    \caption{\textbf{(a)} Skyrmion's trajectory on the plane, \(\textbf{R}(t)\), with initial position \(\textbf{R}(0)=(0,0)\). \textbf{(b)} Limit cycle of \(\dot{\textbf{R}}(t)\) with initial condition \(\dot{\textbf{R}}(0)=(0,0)\). \textbf{(c)} Skyrmion phase space, \(\dot{R}_i(t)\) against \(R_i(t)\), of the horizontal position \(X(t)\) (\(i=x\), blue curve) and vertical position \(Y(t)\) (\(i=y\), red curve). \textbf{(d)} The Skyrmion velocity components \(\dot{R}_i(t)\), with \(i\in\{x,y\}\), plotted against phase, \(\omega t\), for several period of oscillation. This figure shows the transition from the transient state to the steady state, which forms a limit cycle. All of these plots represent Skyrmion dynamics of Neel-type under constant light with \(\eta_1=\eta_2=\alpha\gamma=1\) and \(M_S\omega=40\).}
    \label{fig:constantLight}
\end{figure}

The formulation of total Skyrmion number currents above works for \(Q=1\) cases, which also include meron-pairs with half Skyrmion number each. Since the Skyrmion number current is constructed solely from the topological properties of the texture (i.e., the map from the plane to spherical target space), its existence can also be studied for merons. Furthermore, the equation of motion \eqref{SpatIndependent} is derived directly from the LLG equation, implying that the effect of \(\textbf{J}_Q\) is also predicted for other types of magnetic defects other than Skyrmion. However, since meron-pairs generally do not have an axially symmetric profile, we require a slightly different perturbation scheme which also depends on the radial coordinate. This scheme is beyond the scope of this work. We also note that individual merons are not topologically stable in two dimensions without additional constraints \cite{Manton:2004tk}. Hence, the extension of the framework to other topological textures will be discussed elsewhere.
\section{Skyrmion Trajectory and Limit Cycle}From the BP profile \eqref{BPtypeSol} equipped with the first-order perturbation solution, we can directly calculate all the time-dependent terms in the EOM \eqref{SpatIndependent}, namely \(\mathcal{G}\), \(\textbf{J}_Q\), and \(\textbf{D}\). With these quantities in hand, we can directly solve the EOM to find the Skyrmion's trajectory. To model the trajectory of the Skyrmion, we need to add a phenomenological inertia term that is proportional to \(\ddot{\textbf{R}}\) \cite{saitoh2004current,shiino2017inertia}. This term contains the skyrmion mass, \(M_S\), which arises from fluctuations in the lattice \cite{martinez2017mass,kravchuk2018spin,liu2020measurement,capic2020skyrmion}. In our notation, \(M_S\) has the same dimension as \(t\). After adding this inertial term, the system of equations that is to be solved for the Skyrmion's trajectory is given by
\begin{eqnarray}\label{XEvol}
    M_S\ddot{X}+\mathcal{G}_{xx}\dot{X}+\mathcal{G}_{xy}\dot{Y}+4\pi \left(Q\dot{Y}+ J_{Q,y}\right)&=&D_{x}~,\\
     M_S\ddot{Y}+\mathcal{G}_{xy}\dot{X}+\mathcal{G}_{yy}\dot{Y}-4\pi\left(Q\dot{X}+ J_{Q,x}\right)&=&D_y~.\label{YEvol}
\end{eqnarray}
For the numerical simulations, we chose \(Q=1\) such that \(\textbf{J}_Q\neq0\) in Cartesian basis. We simulate the EOM (\ref{XEvol}-\ref{YEvol}) for the case of constant light with \(B_0>0\).

\begin{figure}
    \centering
    \includegraphics[width=0.5\linewidth]{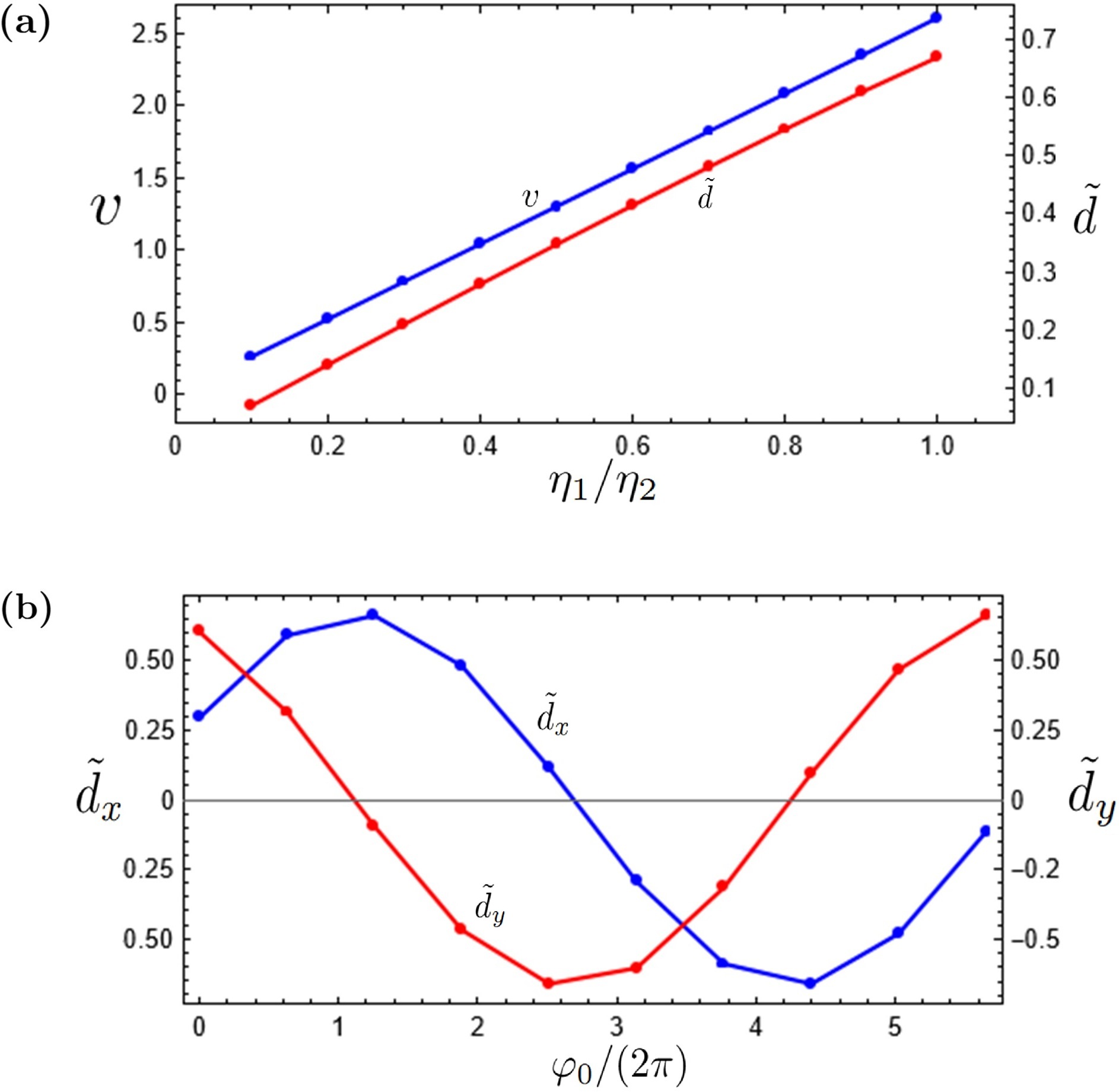}
    \caption{\textbf{(a)} The plot of average magnitude of velocity (\textbf{blue}) and stationary displacement (\textbf{red}) with variation of ratio \(\eta_1/\eta_2\). Both of these quantities are proportionally increasing with respect to the ratio. \textbf{(b)} the direction of displacement represented by the horizontal (\textbf{blue}) and vertical (\textbf{red}) component of \(\tilde{\textbf{d}}\) against various \(\varphi_0\). Each of the data point above are numerically obtained with \((\varphi_0=0,M_S\omega=40,\alpha\gamma=1)\) for figure \textbf{(a)} and \((M_S\omega=40,\eta_1/\eta_2=\alpha\gamma=1)\) for figure \textbf{(b)}.}
    \label{fig:eta_helicity_variation}
\end{figure}
Under a constant circularly polarized light, the EOM defines a limit cycle for \(\dot{\textbf{R}}(t)\). This is deduced from Eq. (\ref{XEvol}-\ref{YEvol}), which belongs to the family of general first-order ordinary differential equations (ODE) in terms of \(\dot{\textbf{R}}\) of the form, \(d\dot{\textbf{R}}/dt+\mathcal{P}(t)\cdot\dot{\textbf{R}}=\textbf{p}(t)\), where \(\mathcal{P}_{ij}(t)\equiv(\mathcal{G}_{ij}+4\pi Q\varepsilon_{ij})/M_S\) and \(\textbf{p}(t)\equiv(\textbf{D}+4\pi\hat{z}\times\textbf{J}_Q)/M_S\). Since the non-oscillatory part of \(\mathcal{P}(t)\) is diagonalizable, the resulting system at large \(t\) is two independent general first-order ODE, such that the late-time behaviour follows the driving term \(\textbf{p}(t)\) and the contribution from the initial value is suppressed by a factor of \(\lim_{t\rightarrow\infty}\left[\exp\left(\int \mathcal{P}\left(t\right)dt\right)\right]^{-1}\). Thus, the evolution of \(\dot{\textbf{R}}\) always converges to a certain limit, regardless of the initial condition, which is the property of limit cycles. This further implies an asymptotically confined \(\textbf{R}(t)\) where confinement region depends on \(\eta_1,~\eta_2\) and \(\alpha\gamma\), as demonstrated in Fig. \ref{fig:constantLight} \textbf{(a)} and \textbf{(b)} for a special case where the initial conditions satisfies \(\textbf{R}(0)=\dot{\textbf{R}}(0)=0\). The phase space trajectory of the Skyrmion is provided in Fig. \ref{fig:constantLight} \textbf{(c)}, where we can see the evolution from the transient state, related to the chosen initial data, which in our case is \(\textbf{R}(0)=\dot{\textbf{R}}(0)=0\), up to the steady state orbit that is reached after a single revolution in momentum space for \(M_S\omega=40\). In general, the phenomenological \(M_S\) is linearly proportional to the relaxation time approaching the limit cycle (See Appendix Sect B2 for proof). The exact evolution of the velocity from the transient to the steady state is demonstrated in Fig. \ref{fig:constantLight} \textbf{(d)} for a certain numerical choice of parameters, and this behaviour is universal up to first-order perturbation. However, it is possible that for more general cases \(Q\geq1\) with higher-order perturbations, behaviour may differ and unstable modes, that do not converge to a limit cycle, might exist.

To study the behaviour of the limit cycle under variations of the physical parameters, we define the average velocity of the Skyrmion as \(v\equiv\sqrt{\langle\dot{X}^2+\dot{Y}^2\rangle_{t}}\) and the stationary displacement as \(\tilde{\textbf{d}}=(\tilde{d}_x,\tilde{d}_y)\equiv\left(\langle X\rangle_t-X(0),\langle Y\rangle_t-Y(0)\right)\), where \(\langle\cdot\rangle_t\) denotes the time-averaged quantity inside the bracket. \(v\) has a geometrical interpretation, namely the effective radius of the limit cycle in momentum space, such that the area of the orbit of \(\dot{\textbf{R}}\) is \(\pi v^2\) (see Appendix Sect. B2 for details). Both of these quantities proportionally increase as \(B_0\) is increased, demonstrated in Fig. \ref{fig:eta_helicity_variation} \textbf{(a)}, which implies that the Skyrmion is displaced further from its initial point for stronger light intensity. However, the confinement area of the Skyrmion at large \(t\) also becomes larger for stronger intensity since it is proportional to \(v\).
Although \(M_S\) determines the evolution at the transient state, it does not affect the displacement of the Skyrmion. Instead, the displacement depends on the \(\varphi_0\), which cannot be controlled by external perturbation \cite{akhir2024stabilization}.

\begin{figure}
    \centering
    \includegraphics[width=0.4\linewidth]{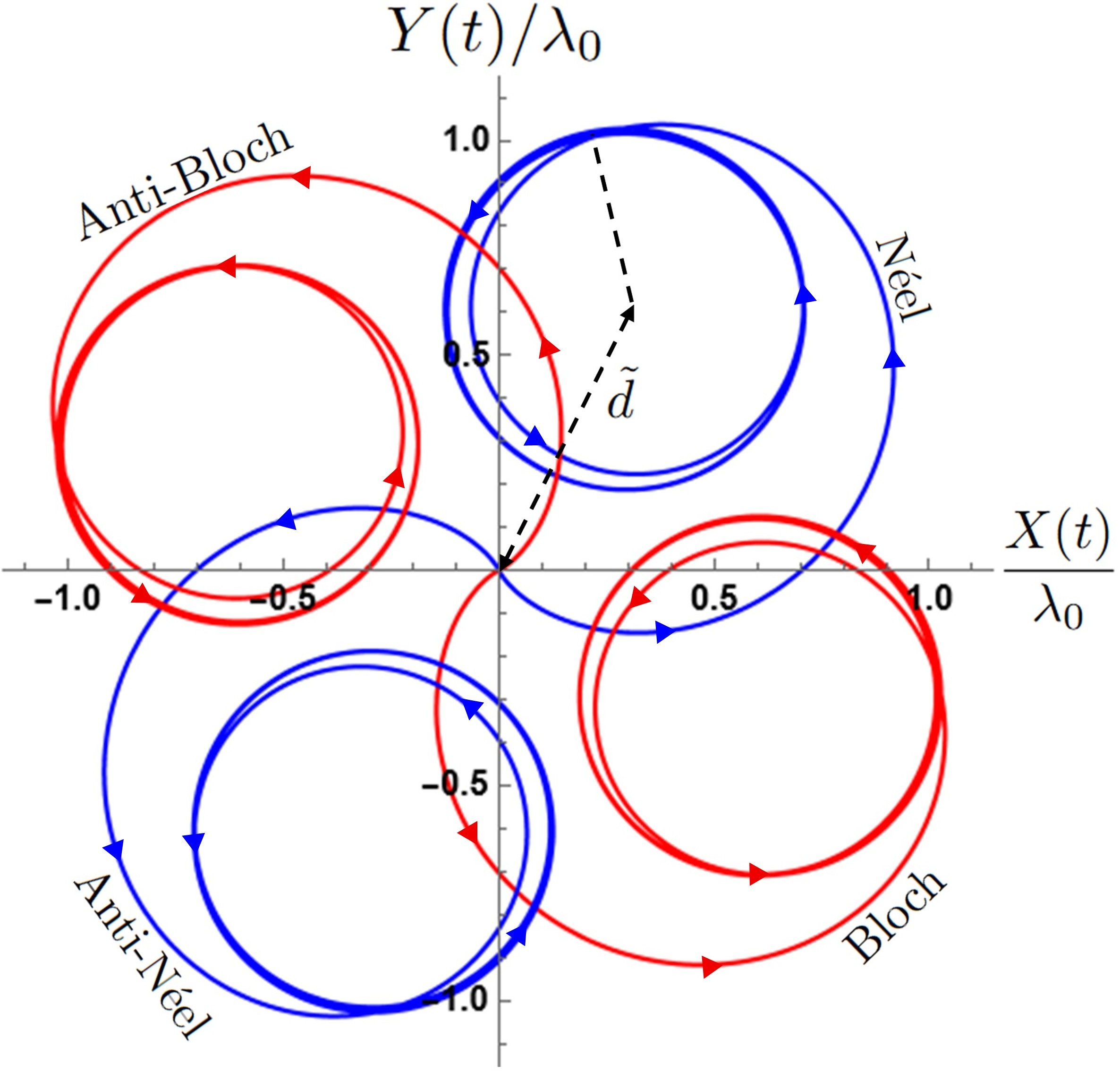}
     \caption{(\textbf{blue} curve) Trajectory of Neel (\(\varphi_0=0\)) and anti-Neel (\(\varphi_0=\pi\)) Skyrmion, and  (\textbf{red} curve) Trajectory of Bloch (\(\varphi_0=\pi/2\)) and anti-Bloch (\(\varphi_0=-\pi/2\)) Skyrmion. We use \(\eta_1/\eta_2=\alpha\gamma=1\) and \(M_S\omega=40\) for the numerical calculation. The displacement magnitude is effectively the distance to the center of the confined orbit \(\textbf{R}\) in real-space.}
    \label{fig:Trajectory_Helicity_Variation}
\end{figure}

In Fig. \ref{fig:eta_helicity_variation} \textbf{(b)} we show the dependence of the components of \(\tilde{\textbf{d}}\) against \(\varphi_0\). This leads to different types of Skyrmions having different trajectories, as shown in Fig. \ref{fig:Trajectory_Helicity_Variation}. Although the magnitude of the displacement is not sensitive to the change in \(\varphi_0\), its direction does depend on \(\varphi_0\). Resulting in a continuous family of confinement regions that is parameterized by \(\varphi_0\). For a group of Skyrmions containing randomly distributed \(\varphi_0\), this family of confinements resembles a ring-like region around their initial position.

\section{Effects of Anisotropy}

As implied by the equation \eqref{PerturbedLLG}, the external magnetic field is treated as a small perturbation, such that we can always decompose the Hamiltonian as follows, 
\begin{equation}\label{HGeneralized}
    \mathcal{H}=\int \left[H_{\text{Sk}}-\textbf{B}_{\text{ext}}(t)\cdot\textbf{n}\right]~d^2x,
\end{equation}
where \(H_{\text{Sk}}\) contains all time-independent terms. Usually, it is expected that \(H_{\text{Sk}}\) contains all necessary terms to stabilize the Skyrmions, since the Skyrmion should persist when the external field is turned off. As such, \(H_\text{Sk}\) determines the profile of the Skyrmion while the external field generates the time-dependency needed for non-zero Skyrmion number current. This implies that the proposed control mechanism is independent of the form of the \(H_{\text{Sk}}\). The minimal model that allows Skyrmion solutions contains only the Heisenberg exchange interaction \cite{Polyakov:1975yp,Arthur:1996ia}. This model is isotropic, giving an axially symmetric Skyrmion profile at the unperturbed limit. The Skyrmion within this minimal model is stabilized at zero temperature through topological protection, where the minimal energy of the system is linear to the topological charge \cite{Fadhilla:2021jiz}. However, this Skyrmion solution does not have a stable radius, according to the Derrick-Hobbart stability conditions \cite{hobart1963instability,Derrick:1964ww}, and it spontaneously disperses into the vacuum state under thermal fluctuations.

The well-known mechanism of Skyrmion stabilization at finite temperature is through the DMI \cite{roessler2006spontaneous}, which arises when we allow anisotropy in the magnetic system \cite{dzyaloshinsky1958thermodynamic}. The anisotropy induces anisotropic exchange interactions, which can be decomposed into symmetric and anti-symmetric parts \cite{moriya1960anisotropic}. The anti-symmetric part of the exchange interaction is the DMI term, which is explicitly given by \(D\textbf{n}\cdot\nabla\times\textbf{n}\), \(D\) is the DMI coupling constant, while the symmetric part is \(J_{ij}\partial_i\textbf{n}\cdot\partial_j\textbf{n}\). \(J_{ij}\) is the symmetric exchange coupling tensor that is related to the Heisenberg coupling through its trace \(J=(J_{xx}+J_{yy})/2\). Crystalline anisotropy can also induce the anisotropy energy, which dictates the preferred direction for the spins \cite{stancil2009spin}. The lowest order anisotropy energy has the form \(K=\left(\textbf{n}_{\text{an}}\cdot\textbf{n}\right)^2\), where \(\textbf{n}_{\text{an}}\) is the preferred direction by the crystalline structure. Thus, a typical model for anisotropic systems has the following \(H_{\text{Sk}}\),
\begin{eqnarray}
    H_{\text{Sk}}^{\text{(an)}}&=&J_{xx}\partial_x\textbf{n}\cdot\partial_x\textbf{n}+J_{yy}\partial_y\textbf{n}\cdot\partial_y\textbf{n}+2J_{xy}\partial_x\textbf{n}\cdot\partial_y\textbf{n}\nonumber\\&&+D\textbf{n}\cdot\nabla\times\textbf{n}-K\left(\textbf{n}_{\text{an}}\cdot\textbf{n}\right)^2.
\end{eqnarray}

Although a full description of Skyrmion motion with anisotropic \(H_{\text{Sk}}\) is beyond the scope of this work, the qualitative description of the case with moderate anisotropy can be studied using the BP solution as the ansatz. For cases with small \(J_{xy}\) compared to \(J\), we can approximate the energy by substituting the BP solution to \eqref{HGeneralized}, resulting in
\begin{eqnarray}
    \mathcal{H}&\simeq& 8\pi J+4\pi D\lambda_0\sin\varphi_0+\pi B_z\left(R^2+2\lambda_0^2\ln\left[\frac{\lambda_0^2}{R^2}\right]\right)\nonumber\\&&-\pi K\left(\cos^2\theta_{\text{an}}R^2+\left(1+3\cos(2\theta_{\text{an}})\right)\lambda_0^2\ln\left[\frac{\lambda_0^2}{R^2}\right]\right).\nonumber\\
\end{eqnarray}
\(R\) is the radius of the system with the assumption that \(R\gg \lambda_0\).
Since in our framework, the circularly polarized light is directed perpendicularly to the plane, then \(B_z=0\). The minimization of \(\mathcal{H}\) with respect to \(\varphi_0\) gives \(\varphi_0=-\text{sgn}[D]~\pi/2\). This implies that the Bloch Skyrmion is the preferred configuration when DMI is present \cite{akhir2024stabilization}. The radius \(\lambda_0\) is also strictly determined by the constants \(D\) and \(K\), according to the minimization of \(\mathcal{H}\) with respect to \(\lambda_0\), which gives the relation \(\lambda_0\left(\ln\left[R^2/\lambda_0^2\right]-1\right)\left(1+3\cos\left(2\theta_{\text{an}}\right)\right)=2|D|/K\). This relation implies that the stable Skyrmion radius comes from the competition between the DMI and crystalline anisotropy energy, which is estimated to be \(\lambda_0\simeq 0.35~|D|/\left(K(1+3\cos(2\theta_{\text{an}}))\right)\), for large \(R\). 

In this moderate anisotropy case, we shall expect that the Heisenberg interaction term is still dominant, which implies that the bound \eqref{BBound} can still be used to estimate \(B_0\). However, the presence of DMI and crystalline anisotropy specifies \(\lambda_0\), as demonstrated above. Thus, we have an upper bound which depends on \(J\), \(D\), and \(K\) simultaneously, as
\begin{equation}
    \omega B_0<1066\frac{\alpha\gamma^2}{1+\alpha^2\gamma^2} \frac{J^2K^4\left(1+3\cos(2\theta_{\text{an}})\right)^4}{D^4}.
\end{equation}
The upper bound is largest at \(\theta_{\text{an}}=0,\pi\) which corresponds to \(\textbf{n}_{\text{an}}=\pm\hat{z}\).
From this estimate, we can conclude qualitatively that to allow higher \(B_0\) we need either larger \(J\) or larger \(K/D\). The larger \(J\) corresponds to the isotropic system. On the other hand, larger \(K/D\) corresponds to a more anisotropic system, where Bloch Skyrmions are more energetically favourable than Neel Skyrmions.

\section{Outlook} Our results indicate that the trajectory of a magnetic Skyrmion with an explicitly time-dependent Hamiltonian can be controlled via its Skyrmion number current. We show that such a current can be generated by an oscillating magnetic field; in particular, we illustrate this using circularly polarized light of constant intensity. The trajectory can be tuned by varying the ratio $\eta_1/\eta_2$, which is achieved by adjusting $B_0$.

According to this theory, we propose that the optical control of magnetic Skyrmions, recently simulated within the micromagnetics framework~\cite{guan2023optically,kazemi2024all} and experimentally demonstrated in magnetic multilayers~\cite{titze2024all}, originates from the dynamics of the Skyrmion number current. This suggests a more direct connection between the topological properties and physically observable quantities of magnetic Skyrmions, beyond the conservation of the Skyrmion number. This hypothesis can be tested experimentally, for example, by observing the motion of a Skyrmionium, a non-topological defect with $Q=0$, under the same circularly polarized light and comparing its trajectory with that of $Q=1$ Skyrmions. The model predicts zero current in the non-topological sector, implying a smaller average velocity than in the $Q=1$ case. A similar test can be performed with Skyrmion bags ($Q>1$), provided the magnetic field strength is sufficiently small to suppress higher-order perturbation effects. Ferrimagnetic \([\text{Fe}/\text{Gd}]\) multilayers are a promising candidate for testing the hypothesis above, as this material has demonstrated all-optical Skyrmion control and Skyrmion breathing mode \cite{titze2024all}. Moreover, being effectively two-dimensional due to its layered structure, it constitutes a planar magnetic system that directly corresponds to the theoretical setup. We also propose \(\text{Cr}\text{I}_3\) for experimental test, because it has been theoretically demonstrated that the material exhibits all-optical Skyrmion control using circularly polarized light \cite{kazemi2024all}.

In line with the earlier proposal by Fujita and Sato~\cite{fujita2017encoding,fujita2017ultrafast}, who simulated the transfer of orbital angular momentum (OAM) to a magnetic system, we employ light carrying spin angular momentum (SAM) with zero OAM and observe a similar effect on the Skyrmion profile. We therefore propose that the driving mechanism of Skyrmion motion arises from the transfer of angular momentum in its most general form, which includes both SAM and OAM. This perspective offers a natural explanation for the distinct trajectories exhibited by Skyrmions with different helicities. To rigorously test this hypothesis, a more realistic system must be considered, as certain interaction symmetries may suppress angular momentum transfer. Addressing this challenge is a key objective for future investigations.

\begin{acknowledgments}
To conclude, E. S. F. acknowledges the support from Badan Riset dan Inovasi Nasional through the Post-Doctoral Program 2024. 
\end{acknowledgments}
\bibliography{main}
\newpage
\appendix
\onecolumngrid

\section{Appendix A: Perturbative Solution of Skyrmion Under Circularly Polarized Light}
As described in the main manuscript, we consider a magnetic system that is influenced by circularly polarized light. We assume the simplest two-dimensional magnetic system where the spins only interact with each other via the Heisenberg exchange interaction, and the light interacts with each spin through the Zeeman effect, resulting in the following form of Hamiltonian,
\begin{equation}\label{Hamiltonian}
    H=J\partial_i\textbf{n}\cdot\partial_i\textbf{n}-\textbf{B}(t)\cdot\textbf{n},
\end{equation}
where \(J\) is the Heisenberg coupling constant and the circularly polarized light, \(B(t)\), is explicitly given by \(\textbf{B}(t)=B_0(\cos(\omega t),\sin(\omega t))=B_0\textbf{p}(t)\). The Zeeman term is then going to be considered as a perturbation term, where the unperturbed Hamiltonian has a steady-state solution satisfying the BP Skyrmion profile. The corresponding LLG equation within this perturbation method is given by \eqref{PerturbedLLG} while the perturbed solution is given by
\begin{equation}
    \textbf{n}=(\sin\Theta\cos\Phi,\sin\Theta\sin\Phi,\cos\Theta),
\end{equation}
where \((\Theta,\Phi)\) is given in BP type form,
\begin{equation}
    \Theta\equiv2\tan^{-1}\left(\frac{r}{\lambda(t,\varphi)}\right),~~~\Phi\equiv\Phi(t,\varphi),
\end{equation}
and the perturbed \((\lambda,\Phi)\) is expanded in terms of perturbation parameter, \(\epsilon\), as
\begin{eqnarray}
    \lambda(t,\varphi)=\lambda_0+\sum_{k=1}^\infty\epsilon^k\tilde{\lambda}_k(t,\varphi),~~~\Phi(t,\varphi)=Q\varphi+\varphi_0+\sum_{k=1}^\infty\epsilon^k\tilde{\Phi}_k(t,\varphi).
\end{eqnarray}
The ansatz \eqref{BPtypeSol} is known as breathing Skyrmion ansatz which can be physically imagined as Skyrmions whose boundary is stretched, in general, anisotropically. at the \(\epsilon=0\) limit, the solution is called the unperturbed solution, denoted by \(\textbf{n}_0\), that is the steady state solution of unperturbed \eqref{PerturbedLLG}, namely \(\textbf{n}_0\times\nabla^2\textbf{n}_0=0\).
\subsection{A1: Energy analysis}
There are constraints for \(\textbf{B}(t)\) in order to have a stable Skyrmion solution using perturbative method. The first one is that \(\textbf{B}(t)\) cannot have out-of-plane component, because this out-of-plane component introduces a more energetically favourable state where all \(\textbf{n}=\hat{z}\), which implies that any Skyrmion on the system will shrink into point with zero radius. With this constraint satisfied, the magnitude \(B_0\) must still be small enough such that the Zeeman effect does not dominate the Heisenberg interaction, since the unperturbed \(\textbf{n}_0\) is induced by this interaction. We can find the upper bound for \(B_0\) from the comparison between the energy of light and the energy dissipated by Gilbert's damping. Let \(U=\int H~d^2x\), be the total energy of the system, hence the rate of the energy over time is \(dU/dt=\int\left[\left(\delta H/\delta\textbf{n}\right)\cdot\partial_t\textbf{n}+\partial_tH\right]d^2x\). We know from the LLG equation that \(\partial_t\textbf{n}=\gamma\left[\textbf{n}\times(\delta H/\delta\textbf{n})+\alpha\gamma\textbf{n}\times\left(\textbf{n}\times(\delta H/\delta\textbf{n})\right)\right]/(1+\alpha^2\gamma^2)\). Thus, we arrive at 
\begin{equation}\label{EnergyRate}
    \frac{dU}{dt}=\int\left[\frac
    {\alpha\gamma^2}{1+\alpha^2\gamma^2}\left(\left(\frac{\delta H}{\delta\textbf{n}}\cdot\textbf{n}\right)^2-\frac{\delta H}{\delta\textbf{n}}\cdot\frac{\delta H}{\delta\textbf{n}}\right)+\partial_tH\right]d^2x~.
\end{equation} This energy rate must be bounded above by zero, \(dU/dt\leq0\), which physically implies that the dissipated energy is always higher than the energy supplied by the light.
We need to estimate the bound above for a single period of light. One possible way to satisfy the bound for \(dU/dt\) is by considering a stronger bound where the integrand in \eqref{EnergyRate} is bounded by zero, namely
\begin{equation}
    \max_{t\in[0,2\pi/\omega]}\left(\frac{1+\alpha^2\gamma^2}{\alpha\gamma^2}\partial_tH-\frac{\delta H}{\delta\textbf{n}}\cdot\frac{\delta H}{\delta\textbf{n}}\right)<0~,
\end{equation}
where we have dropped the first term since it is a positive-definite term. Now, for our case with the Hamiltonian \eqref{Hamiltonian}, we have an effective magnetic field \(-\delta H/\delta \textbf{n}=2J\nabla^2\textbf{n}+\textbf{B}(t)\). As such, we can recast the inequality above as
\begin{eqnarray}
    &&\frac{1+\alpha^2\gamma^2}{\alpha\gamma^2}\max_{t\in[0,2\pi/\omega]}\partial_tH<\min_{t\in[0,2\pi/\omega]}\frac{\delta H}{\delta\textbf{n}}\cdot\frac{\delta H}{\delta\textbf{n}}\nonumber\\
    &\Rightarrow& \frac{1+\alpha^2\gamma^2}{\alpha\gamma^2}\max_{t\in[0,2\pi/\omega]}\left(-\partial_t\textbf{B}(t)\cdot\textbf{n}\right)\nonumber\\&&<\min_{t\in[0,2\pi/\omega]}\|2J\nabla^2\textbf{n}+\textbf{B}(t)\|.
\end{eqnarray}
 This inequality can be estimated using the unperturbed Skyrmion texture that solves \(\textbf{n}_0\times\nabla^2\textbf{n}_0=0\) such that
 \begin{eqnarray}\label{ineqSpaceDependent}
     &&\frac{1+\alpha^2\gamma^2}{\alpha\gamma^2}B_0\max_{t\in[0,2\pi/\omega]}\left(-\partial_t\textbf{p}(t)\cdot\textbf{n}_0\right) <\|2J\nabla^2\textbf{n}_0\|^2.
 \end{eqnarray}
The unperturbed solution \(\textbf{n}_0\) is well-known as the Belavin-Polyakov (BP) solution, that is given by \(\textbf{n}_0=(2\lambda_0(x\cos\varphi_0-y\sin\varphi_0),~2\lambda_0(x\sin\varphi_0+y\cos\varphi_0),~\lambda_0^2-r^2)/(\lambda_0^2+r^2)\), where \(r^2=x^2+y^2\), \(\lambda_0\) is the Skyrmion radius, and \(\varphi_0\) is the unperturbed Skyrmion helicity.
 The LHS of \eqref{ineqSpaceDependent} is \((x,y)\) dependent through \(\textbf{p}\cdot\textbf{n}_0\) that is maximized at \(r=\lambda_0\). Let \(\nu\) be the phase \(\nu=\omega t\) such that \(\textbf{p}=(\cos\nu,\sin \nu)\), then at \(r=\lambda_0\) we have \(\textbf{p}\cdot\textbf{n}_0=\cos(\varphi+\nu+\varphi_0)\) where \(\varphi=\tan^{-1}(y/x)\). As such, \(-\partial_t\textbf{p}\cdot\textbf{n}_0=\omega\sin(\varphi+\nu+\varphi_0)\) whose maximum value is equal to one. On the other hand, the RHS of \eqref{ineqSpaceDependent} is \((x,y)\) dependent through \(\|\nabla^2\textbf{n}_0\|^2\). If we substitute the BP solution into this norm, we have \(\|\nabla^2\textbf{n}_0\|^2=64\lambda^4/(\lambda_0^2+r^2)^4\) that is evaluated to \(4/\lambda_0^4\) at \(r=\lambda_0\). As a result, the estimated upper bound for \(B_0\) is given by
 \begin{equation}\label{ineqB0}
      \omega B_0 <\frac{\alpha\gamma^2}{1+\alpha^2\gamma^2}\frac{16J^2}{\lambda_0^4}.
 \end{equation}
Beyond this value of \(B_0\), the Zeeman effect will become significant compared to the exchange interaction and the topological stability of the solution \(\textbf{n}(t,\textbf{r})\) is not guaranteed. 

\subsection{A2: Profile of Skyrmion texture}
Up to first order of \(\epsilon\), the equations for \(\tilde{\lambda}_1\) and \(\tilde{\Phi}_1\) from evaluating \eqref{PerturbedLLG} at the boundary of the Skyrmion are
\begin{eqnarray}\label{FirstPerturb}
    \partial_t\tilde{\Phi}_1+\alpha\gamma\frac{\partial_t\tilde{\lambda}_1}{\lambda_0}-2\frac{\gamma J}{\lambda_0^2}\frac{\partial_\varphi^2\tilde{\lambda}_1}{\lambda_0}+2(1-Q^2)\frac{\gamma J}{\lambda_0^2}\frac{\tilde{\lambda}_1}{\lambda_0}&=&0\\
    \frac{\partial_t\tilde{\lambda}_1}{\lambda_0}-\gamma B_0\sin\left(Q\varphi+\varphi_0-\omega t\right)-\alpha\gamma\partial_t\tilde{\Phi}_1+2\frac{\gamma J}{\lambda_0^2}\partial_\varphi^2\tilde{\Phi}_1&=&0.\label{SecPerturb}
\end{eqnarray}
Both \(\tilde{\lambda}_1\) and \(\tilde{\Phi}_1\) must be periodic for \(\varphi\rightarrow\varphi+2\pi\). The homogenous part of (\ref{FirstPerturb}-\ref{SecPerturb}) is diffusive, hence the homogenous solutions are suppressed at the steady state limit. At steady state, the non-homogenous solution should behave like the driving term controlled by \(B_0\). As such, the solutions are \(\tilde{\lambda}_1/\lambda_0=A^1_\lambda\cos\left(Q\varphi+\varphi_0-\omega t\right)+A^2_\lambda\sin\left(Q\varphi+\varphi_0-\omega t\right)\) and \(\tilde{\Phi}_1=A^1_\Phi\cos\left(Q\varphi+\varphi_0-\omega t\right)+A^2_\Phi\sin\left(Q\varphi+\varphi_0-\omega t\right)\). Substituting these solutions to \eqref{FirstPerturb} and \eqref{SecPerturb} gives us
\begin{eqnarray}
    0&=&\left[\omega A^1_{\Phi}+\alpha\gamma \omega A^1_{\lambda}+2\frac{\gamma J}{\lambda_0^2}A^2_{\lambda}\right]\sin\left(Q\varphi+\varphi_0-\omega t\right)\nonumber\\
    &&+\left[-\omega A^2_{\Phi}-\alpha\gamma\omega A^2_{\lambda}+2\frac{\gamma J}{\lambda_0^2}A^1_{\lambda}\right]\cos\left(Q\varphi+\varphi_0-\omega t\right),\\
    0&=&\left[\omega A^1_{\lambda}-\gamma B_0-\alpha\gamma\omega A^1_{\Phi}-2\frac{\gamma J}{\lambda_0^2}Q^2A^2_{\Phi}\right]\sin\left(Q\varphi+\varphi_0-\omega t\right)\nonumber\\
    &&+\left[-\omega A^2_{\lambda}+\alpha\gamma\omega A^2_{\Phi}-2\frac{\gamma J}{\lambda_0^2}Q^2A^1_{\Phi}\right]\cos\left(Q\varphi+\varphi_0-\omega t\right).
\end{eqnarray}
The equation above must be satisfied for all \(t\). Thus, we can solve each coefficient for \((A^1_{\Phi},A^2_{\Phi},A^1_{\lambda},A^2_{\lambda})\). The solutions for these \(A^i_\mu\)s is then can be substituted back to \(\lambda(t,\varphi)\) and \(\Phi(t,\varphi)\), such that their solutions, up to first order perturbation, are given by
\begin{eqnarray}
    \lambda(t,\varphi)&=&\lambda_0+\lambda_0\frac{\eta_1\left(1+\alpha^2\gamma^2-\eta_2^2Q^2\right)}{\alpha^4\gamma^4+\left(\eta_2^2Q^2-1\right)^2+\alpha^2\gamma^2\left(\eta_2^2(1+Q^4)+2\right)}\cos\left(Q\varphi+\varphi_0-\omega t\right)\nonumber\\&&+\lambda_0\frac{\alpha\gamma\eta_1\eta_2\left(1+Q^2\right)}{\alpha^4\gamma^4+\left(\eta_2^2Q^2-1\right)^2+\alpha^2\gamma^2\left(\eta_2^2(1+Q^4)+2\right)}\sin\left(Q\varphi+\varphi_0-\omega t\right)~,\\
    \Phi(t,\varphi)&=&Q\varphi+\varphi_0-\frac{\alpha\gamma\eta_1\left(1+\alpha^2\gamma^2+\eta_2^2\right)}{\alpha^4\gamma^4+\left(\eta_2^2Q^2-1\right)^2+\alpha^2\gamma^2\left(\eta_2^2(1+Q^4)+2\right)}\cos\left(Q\varphi+\varphi_0-\omega t\right)\nonumber\\
    &&+\frac{\eta_1\eta_2\left(1-\left(\alpha^2\gamma^2+\eta_2^2\right)Q^2\right)}{\alpha^4\gamma^4+\left(\eta_2^2Q^2-1\right)^2+\alpha^2\gamma^2\left(\eta_2^2(1+Q^4)+2\right)}\sin\left(Q\varphi+\varphi_0-\omega t\right)~,
\end{eqnarray}
where we have defined two physical parameters of the system, namely \(\eta_1\) and \(\eta_2\). The ratio \(\eta_1/\eta_2=B_0\lambda_0^2/(2J)\), is the measure of the external magnetic field's strength compared to the magnetic interaction between the spins. The upper bound for \(B_0\), can be recast into relation between \(\eta_1\) and \(\eta_2\) as described in the main manuscript.
\subsection{A3: Skyrmion number current, shape factor, and driving term}
Since the Skyrmion profile depends explicitly on time, as shown in  \eqref{lambdat} and \eqref{phit}, the Skyrmion number current is non-zero. For the BP Skyrmion, the corresponding current is given by
\begin{equation}
    \textbf{j}=\frac{\lambda}{\pi(\lambda^2+r^2)^2}\begin{bmatrix}
        x\left(\partial_t\lambda\partial_\varphi\Phi-\partial_t\Phi\partial_\varphi\lambda\right)+y\lambda\partial_t\Phi\\
        y\left(\partial_t\lambda\partial_\varphi\Phi-\partial_t\Phi\partial_\varphi\lambda\right)-x\lambda\partial_t\Phi
    \end{bmatrix},
\end{equation}
and the \(\partial_t\textbf{n}\cdot\partial_i\textbf{n}\) terms are given by
\begin{equation}
    \begin{bmatrix}
        \partial_t\textbf{n}\cdot\partial_x\textbf{n}\\\partial_t\textbf{n}\cdot\partial_y\textbf{n}
    \end{bmatrix}=\frac{4}{\left(\lambda^2+r^2\right)^2}\begin{bmatrix}
        -y\left(\lambda^2\partial_t\Phi\partial_\varphi\Phi+\partial_t\lambda\partial_\varphi\lambda\right)-x\lambda\partial_t\lambda\\
        x\left(\lambda^2\partial_t\Phi\partial_\varphi\Phi+\partial_t\lambda\partial_\varphi\lambda\right)-y\lambda\partial_t\lambda
    \end{bmatrix}~.
\end{equation}
The total Skyrmion number current \(\textbf{J}_{Q}\) is then given by the integral
\begin{equation}\label{TotCurrent}
    \textbf{J}_Q=\frac{1}{4}\int_0^{2\pi}\begin{bmatrix}
        \cos\varphi\left(\partial_t\lambda\partial_\varphi\Phi-\partial_t\Phi\partial_\varphi\lambda\right)+\sin\varphi\lambda\partial_t\Phi\\
        \sin\varphi\left(\partial_t\lambda\partial_\varphi\Phi-\partial_t\Phi\partial_\varphi\lambda\right)-\cos\varphi\lambda\partial_t\Phi
    \end{bmatrix}d\varphi.
\end{equation}
Let us also define \(D_i(t)\equiv\gamma\alpha\int\partial_t\textbf{n}\cdot\partial_i\textbf{n}~d^2x\), such that
\begin{equation}
    \textbf{D}=\gamma\alpha\int_0^{2\pi}\frac{\pi}{\lambda}\begin{bmatrix}
        -\sin\varphi\left(\lambda^2\partial_t\Phi\partial_\varphi\Phi+\partial_t\lambda\partial_\varphi\lambda\right)-\cos\varphi\lambda\partial_t\lambda\\
        \cos\varphi\left(\lambda^2\partial_t\Phi\partial_\varphi\Phi+\partial_t\lambda\partial_\varphi\lambda\right)-\sin\varphi\lambda\partial_t\lambda
    \end{bmatrix}d\varphi
\end{equation}
We can see that even if both \(\lambda\) and \(\Phi\) depend on \(t\) but do not depend on \(\varphi\) then \(\textbf{J}_Q=0\). In other words, axial symmetry breaking is necessary to drive the Skyrmion's motion using Skyrmion number density. This implies that axially symmetric Skyrmions are not sensitive to the Skyrmion number current in their vicinity.

For the perturbative steady state solutions \eqref{lambdat} and \eqref{phit}, the integral \eqref{TotCurrent} yields non zero value, if and only if \(Q=1\). Thus, we are going to focus on the case with unit Skyrmion number from this point on. The remaining non-zero terms for this \(Q=1\) case are,
\begin{eqnarray}
    \int_0^{2\pi}\sin\varphi\lambda\partial_t\Phi~d\varphi&=&\pi\lambda_0\omega\left(A^1_\Phi\cos\left(\varphi_0-\omega t\right)+A^2_\Phi\sin\left(\varphi_0-\omega t\right)\right),\\
    \int_0^{2\pi}\cos\varphi\lambda\partial_t\Phi~d\varphi&=&\pi\lambda_0\omega\left(A^1_\Phi\sin\left(\varphi_0-\omega t\right)-A^2_\Phi\cos\left(\varphi_0-\omega t\right)\right),\\
    \int_0^{2\pi}\cos\varphi\partial_t\lambda\partial_\varphi\Phi~d\varphi&=&\pi\lambda_0\omega\left(A^1_\lambda\sin\left(\varphi_0-\omega t\right)-A^2_\lambda\cos\left(\varphi_0-\omega t\right)\right),\\
    \int_0^{2\pi}\sin\varphi\partial_t\lambda\partial_\varphi\Phi~d\varphi&=&\pi\lambda_0\omega\left(A^1_\lambda\cos\left(\varphi_0-\omega t\right)+A^2_\lambda\sin\left(\varphi_0-\omega t\right)\right).
\end{eqnarray}
Substituting this calculation to the expression \eqref{TotCurrent}, we have
\begin{equation}
    \textbf{J}_Q=\begin{bmatrix}
        J_{Q}^-\\
        J_{Q}^+
    \end{bmatrix}\cos\left(\varphi_0-\omega t\right)+\begin{bmatrix}
        J_{Q}^+\\
        -J_{Q}^-
    \end{bmatrix}\sin\left(\varphi_0-\omega t\right),
\end{equation}
where we have defined a shorthand notation,
\begin{eqnarray}
    J_Q^-&\equiv&\frac{\pi\lambda_0\omega}{4}\left[A^1_\Phi-A^2_\lambda\right],\\
    J_Q^+&\equiv&\frac{\pi\lambda_0\omega}{4}\left[A^1_\lambda+A^2_\Phi\right].
\end{eqnarray}
Following a similar procedure, we need the following calculations to calculate \(\textbf{D}\),
\begin{eqnarray}
    \int_0^{2\pi}\sin\varphi\lambda\partial_t\Phi\partial_\varphi\Phi~d\varphi&=&-\frac
    {\pi\lambda_0\omega}{4}\left[-\left(2A_\lambda^1A^1_\Phi  A^2_\Phi-A^2_\lambda\left(A^2_\Phi{}^2+3A^1_\Phi{}^2\right)\right)\cos\left(\varphi_0-\omega t\right)\right.\nonumber\\&&\left.+\left(2A_\lambda^2A^1_\Phi  A^2_\Phi-A^1_\lambda\left(A^1_\Phi{}^2+3A^2_\Phi{}^2\right)\right)\sin\left(\varphi_0-\omega t\right)\right],\\
    \int_0^{2\pi}\cos\varphi\partial_t\lambda~d\varphi&=&\pi\lambda_0\omega\left(A^1_\lambda\sin\left(\varphi_0-\omega t\right)-A^2_\lambda\cos\left(\varphi_0-\omega t\right)\right)~,\\
    \int_0^{2\pi}\cos\varphi\lambda\partial_t\Phi\partial_\varphi\Phi~d\varphi&=&\frac
    {\pi\lambda_0\omega}{4}\left[\left(2A_\lambda^2A^1_\Phi  A^2_\Phi-A^1_\lambda\left(A^1_\Phi{}^2+3A^2_\Phi{}^2\right)\right)\cos\left(\varphi_0-\omega t\right)\right.\nonumber\\&&\left.+\left(2A_\lambda^1A^1_\Phi  A^2_\Phi-A^2_\lambda\left(A^2_\Phi{}^2+3A^1_\Phi{}^2\right)\right)\sin\left(\varphi_0-\omega t\right)\right],\\
    \int_0^{2\pi}\sin\varphi\partial_t\lambda~d\varphi&=&\pi\lambda_0\omega\left(A^1_\lambda\cos\left(\varphi_0-\omega t\right)+A^2_\lambda\sin\left(\varphi_0-\omega t\right)\right).
\end{eqnarray}
Using these integrals, we arrive at the expression for \(\textbf{D}\) as follows
\begin{equation}
    \textbf{D}=\begin{bmatrix}
        D^-\\D^+
    \end{bmatrix}\cos\left(\varphi_0-\omega t\right)+\begin{bmatrix}
        D^+\\-D^-
    \end{bmatrix}\sin\left(\varphi_0-\omega t\right),
\end{equation}
where we have simplify the notations using
\begin{eqnarray}
    D^-&\equiv&\frac{\alpha\gamma\pi^2\lambda_0\omega}{4}\left[-\left(2A_\lambda^1A^1_\Phi  A^2_\Phi-A^2_\lambda\left(A^2_\Phi{}^2+3A^1_\Phi{}^2\right)\right)+4A^2_\lambda\right],\\
     D^+&\equiv&\frac{\alpha\gamma\pi^2\lambda_0\omega}{4}\left[\left(2A_\lambda^2A^1_\Phi  A^2_\Phi-A^1_\lambda\left(A^1_\Phi{}^2+3A^2_\Phi{}^2\right)\right)-4A^1_\lambda\right].
\end{eqnarray}

The disturbance from the external magnetic field does not just generate the Skyrmion number current and driving term \(\textbf{D}\), but also induces deformation in the Skyrmion shape. This deformation can be studied through the strain tensor. Substituting the solution \eqref{BPtypeSol} to the expression of the components of \(\mathcal{D}\) gives us
\begin{eqnarray}
    \mathcal{D}_{xx}&=&4\frac{\left(x\lambda+y\partial_\varphi\lambda\right)^2+y^2\lambda^2(\partial_\varphi\Phi)^2}{r^2\left(r^2+\lambda^2\right)^2}~,\\
    \mathcal{D}_{yy}&=&4\frac{\left(y\lambda-x\partial_\varphi\lambda\right)^2+x^2\lambda^2(\partial_\varphi\Phi)^2}{r^2\left(r^2+\lambda^2\right)^2}~,\\
    \mathcal{D}_{xy}&=&-4\frac{\left(x^2-y^2\right)\lambda\partial_\varphi\lambda+xy\left(\partial_\varphi\lambda\right)^2+xy\lambda^2\left((\partial_\varphi\Phi)^2-1\right)}{r^2\left(r^2+\lambda^2\right)^2}~.
\end{eqnarray}
We can see that for axially symmetric Skyrmion, where \(\lambda=\lambda_0\) and \(\Phi=\varphi+\varphi_0\), \(\mathcal{D}_{xy}=0\) and \(\mathcal{D}_{xx}=\mathcal{D}_{yy}\).
The shape factor of the Skyrmion can be found by integrating the above strain tensor over the whole plane. The radial integrals of the strain tensor is straight forward, which leaves us with the following angular integrals for the shape factor
\begin{eqnarray}
    \mathcal{G}_{xx}&=&2\alpha\gamma\int_0^{2\pi}\frac{\left(\cos\varphi\lambda+\sin\varphi\partial_{\varphi}\lambda\right)^2+\lambda^2(\partial_\varphi\Phi)^2\sin^2\varphi}{\lambda^2}~d\varphi~,\\
    \mathcal{G}_{yy}&=&2\alpha\gamma\int_0^{2\pi}\frac{\left(\sin\varphi\lambda-\cos\varphi\partial_{\varphi}\lambda\right)^2+\lambda^2(\partial_\varphi\Phi)^2\cos^2\varphi}{\lambda^2}~d\varphi~,\\
    \mathcal{G}_{xy}&=&-\alpha\gamma\int_0^{2\pi}\frac{2\cos\left(2\varphi\right)\lambda\partial_\varphi\lambda+\sin\left(2\varphi\right)\left(\partial_\varphi\lambda\right)^2+\sin\left(2\varphi\right)\lambda^2\left((\partial_\varphi\Phi)^2-1\right)}{\lambda^2}~d\varphi~.
\end{eqnarray}
Substituting the first-order perturbations in \eqref{lambdat} and \eqref{phit}, into the expressions of \(\mathcal{G}_{ij}\) above and taking only the leading order of the perturbation gives us \(\mathcal{G}_{xy}=0\) and 
\begin{eqnarray}
    \frac{\mathcal{G}_{xx}}{\alpha\gamma\pi}
    &=&\left[4+\left((A^1_\Phi)^2+(A^2_\Phi)^2+((A^1_\lambda)^2+(A^2_\lambda)^2\right)\right]\\
    &&+\frac{(A^1_\Phi)^2-(A^2_\Phi)^2+(A^1_\lambda)^2-(A^2_\lambda)^2}{2}\cos\left(2(\varphi_0-\omega t)\right)\nonumber\\&&+\left(A^1_\Phi A^2_\Phi+A^1_\lambda A^2_\lambda\right)\sin\left(2(\varphi_0-\omega t)\right)~,\nonumber\\
    \frac{\mathcal{G}_{yy}}{\alpha\gamma\pi}
    &=&\left[4+\left((A^1_\Phi)^2+(A^2_\Phi)^2+((A^1_\lambda)^2+(A^2_\lambda)^2\right)\right]\\
    &&+\frac{(A^2_\Phi)^2-(A^1_\Phi)^2+(A^2_\lambda)^2-(A^1_\lambda)^2}{2}\cos\left(2(\varphi_0-\omega t)\right)\nonumber\\&&-\left(A^1_\Phi A^2_\Phi+A^1_\lambda A^2_\lambda\right)\sin\left(2(\varphi_0-\omega t)\right)~.\nonumber
\end{eqnarray}
The resulting shape factor of the Skyrmion can be written as the following expression
\begin{equation}
    \mathcal{G}=\begin{bmatrix}
        \mathcal{G}^0&0\\0&\mathcal{G}^0
    \end{bmatrix}+\begin{bmatrix}
        \mathcal{G}^+&0\\0&-\mathcal{G}^+
    \end{bmatrix}\cos\left(2(\varphi_0-\omega t)\right)+\begin{bmatrix}
        \mathcal{G}^-&0\\0&-\mathcal{G}^-
    \end{bmatrix}\sin\left(2(\varphi_0-\omega t)\right),
\end{equation}
where \(\mathcal{G}^0,~\mathcal{G}^+\) and \(\mathcal{G}^-\) are given by
\begin{eqnarray}
    \mathcal{G}^0&=&\alpha\gamma\pi\left[4+\left((A^1_\Phi)^2+(A^2_\Phi)^2+((A^1_\lambda)^2+(A^2_\lambda)^2\right)\right]~,\\
    \mathcal{G}^+&=&\alpha\gamma\pi\frac{(A^1_\Phi)^2-(A^2_\Phi)^2+(A^1_\lambda)^2-(A^2_\lambda)^2}{2}~,\\
    \mathcal{G}^-&=&\alpha\gamma\pi\left(A^1_\Phi A^2_\Phi+A^1_\lambda A^2_\lambda\right)~.
\end{eqnarray}

Together with \(\textbf{D}\) and \(\textbf{J}_Q\), these explicit expressions are used in the numerical simulations of Skyrmion trajectory to solve for \(\textbf{R}(t)\).

\section{Appendix B: Skyrmion Trajectory}
The Skyrmion trajectory is found by solving \eqref{XEvol} and \eqref{YEvol} where the components of \(\mathcal{G},~\textbf{D}\) and \(\textbf{J}_Q\) are substituted from the results of the previous section.

\subsection{B1: The limit cycle}
The system of two equations above can be recast as an ODE
\begin{equation}
    \frac{d\dot{\textbf{R}}}{dt}+\mathcal{P}(t)\cdot\dot{\textbf{R}}=\textbf{p}(t),
\end{equation}
where \(P\) and \(\textbf{p}\) is given by
\begin{equation}
    \mathcal{P}=\frac{1}{M_S}\begin{bmatrix}
          \mathcal{G}_{xx}&\mathcal{G}_{xy}+4\pi Q\\
        \mathcal{G}_{xy}-4\pi Q&\mathcal{G}_{yy}
    \end{bmatrix},~\textbf{p}=\frac{1}{M_S}\begin{bmatrix}
        D_x-4\pi J_{Q,y}\\
        D_y+4\pi J_{Q,x}
    \end{bmatrix}.
\end{equation}
For our case with \(Q=1\) Skyrmion up to first-order perturbation, we have
\begin{eqnarray}
    \mathcal{P}&=&\frac{1}{M_S}\begin{bmatrix}
          \mathcal{G}^0&4\pi \\
        -4\pi &\mathcal{G}^0
    \end{bmatrix}+\frac{\mathcal{G}^+\cos\left(2\left(\varphi_0-\omega t\right)\right)+\mathcal{G}^-\sin\left(2\left(\varphi_0-\omega t\right)\right)}{M_S}\begin{bmatrix}
        1&0\\
        0&-1
    \end{bmatrix}~,\\
    \textbf{p}&=&\frac{1}{M_S}\begin{bmatrix}
        \left(D^--4\pi J_{Q}^+\right)\cos\left(\varphi_0-\omega t\right)+\left(D^++4\pi J_Q^-\right)\sin\left(\varphi_0-\omega t\right)\\
         \left(D^++4\pi J_Q^-\right)\cos\left(\varphi_0-\omega t\right)-\left(D^--4\pi J_{Q}^+\right)\sin\left(\varphi_0-\omega t\right)
    \end{bmatrix}~.
\end{eqnarray}
The important quantity in solving the above equation is the integration factor \(\exp{}\left(\int^t \mathcal{P}(\tilde{t}) d\tilde{t}\right)\) that contains
\begin{equation}
    \int^t \mathcal{P}(\tilde{t}) d\tilde{t}=\frac{t}{M_S}\begin{bmatrix}
          \mathcal{G}^0&4\pi \\
        -4\pi &\mathcal{G}^0
    \end{bmatrix}+\frac{\mathcal{G}^-\cos\left(2\left(\varphi_0-\omega t\right)\right)-\mathcal{G}^+\sin\left(2\left(\varphi_0-\omega t\right)\right)}{2\omega M_S}\begin{bmatrix}
        1&0\\
        0&-1
    \end{bmatrix}~,
\end{equation}
which goes like
\begin{equation}
    \int^t \mathcal{P}(\tilde{t}) d\tilde{t}\approx\frac{t}{M_S}\begin{bmatrix}
          \mathcal{G}^0&4\pi \\
        -4\pi &\mathcal{G}^0
    \end{bmatrix},~~~~\text{at large}~t.
\end{equation}
We can see that the oscillatory part of \(\mathcal{G}\) does not contribute to teh integration factor at large \(t\). Thus, our system of equation is effectively becomes
    \begin{equation}
    \frac{d\dot{\textbf{R}}}{dt}+\begin{bmatrix}
          \mathcal{G}^0&4\pi \\
        -4\pi &\mathcal{G}^0
    \end{bmatrix}\cdot\frac{\dot{\textbf{R}}}{M_S}\approx\frac{1}{M_S}\begin{bmatrix}
        \left(D^--4\pi J_{Q}^+\right)\cos\left(\varphi_0-\omega t\right)+\left(D^++4\pi J_Q^-\right)\sin\left(\varphi_0-\omega t\right)\\
         \left(D^++4\pi J_Q^-\right)\cos\left(\varphi_0-\omega t\right)-\left(D^--4\pi J_{Q}^+\right)\sin\left(\varphi_0-\omega t\right)
    \end{bmatrix},
\end{equation}
at large \(t\), which is diagonalizable into two independent one-dimensional general ODE. The initial value contribution in the solutions of one-dimensional general ODE is suppressed exponentially at large \(t\), leaving only the driving term. This implies that the stationary solutions of our system also does not depend on the initial values, hence a limit cycle.

For this specific case, diagonalization includes complexification of \((X,Y)\) into \(\left((Y-iX),(Y+iX)\right)\) with complex eigenvalues of \(\mathcal{P}\). Although the solutions of \((X,Y)\) are still \(\mathbb{R}^2\)-valued, solving the diagonalized equation must be done on \(\mathbb{C}\). However, an alternative approach with a completely \(\mathbb{R}^2\)-valued vector is available. Since, at large \(t\), \(\mathcal{P}\) commutes with \(\int^t\mathcal{P}(\tilde{t})d\tilde{t}\), if we expand the integration factor at large \(t\) as
\begin{equation}
    e^{\int^t\mathcal{P}(\tilde{t})d\tilde{t}}\approx\mathbb{I}+\int^t\mathcal{P}_\infty d\tilde{t}+\frac{1}{2!}\left(\int^t\mathcal{P}_\infty d\tilde{t}\right)^2+\dots,
\end{equation}
where \(\mathbb{I}\) is the identity matrix, then we have the following relation
\begin{eqnarray}
    \frac{d}{dt}e^{\int^t\mathcal{P}(\tilde{t})d\tilde{t}}&\approx&\mathcal{P}_\infty +\mathcal{P}_\infty \left(\int^t\mathcal{P}_\infty d\tilde{t}\right)+\frac{1}{2!}\mathcal{P}_\infty \left(\int^t\mathcal{P}_\infty d\tilde{t}\right)^2\dots\nonumber\\
    &=&\mathcal{P}_\infty e^{\int^t\mathcal{P}_\infty ~d\tilde{t}}
\end{eqnarray}
Here, \(\mathcal{P}_\infty\) denotes the non-oscillatory part of \(\mathcal{P}(t)\) that is not suppressed at large \(t\). We can use the above property to recast the ODE as
\begin{equation}\label{LargeTODE}
    \frac{d}{dt}\left[e^{\int^t\mathcal{P}_\infty ~d\tilde{t}}\cdot\dot{\textbf{R}}\right]\approx e^{\int^t\mathcal{P}_\infty ~d\tilde{t}}\cdot\textbf{p}(t).
\end{equation}
Equation \eqref{LargeTODE} has a general solution of the form
\begin{equation}
    \dot{\textbf{R}}(t)\approx\left(e^{\int^t\mathcal{P}_\infty~d\tilde{t}}\right)^{-1}\int_0^te^{\int^{t'}\mathcal{P}_\infty~d\tilde{t}}\cdot\textbf{p}(t')~dt'+\left(e^{\int^t\mathcal{P}_\infty~d\tilde{t}}\right)^{-1}\cdot\dot{\textbf{R}}(0).
\end{equation}
Now, we need to check the behaviour of the integration factor at large \(t\). Notice that
\begin{equation}
    \int^t\mathcal{P}_\infty~d\tilde{t}=\frac{t}{M_S}\left[\mathcal{G}^0\mathbb{I}+4\pi \sigma\right],~\text{with}~\sigma=\begin{bmatrix}
        0&-1\\1&0
    \end{bmatrix}~.
\end{equation}
Since the commutator \([\mathbb{I},\sigma]=0\), using the Baker-Campbell-Hausdorff formula, we can write the large \(t\) integration factor as
\begin{equation}
    e^{\int^t\mathcal{P}_\infty~d\tilde{t}}=e^{\frac{\mathcal{G}^0t}{M_S}\mathbb{I}}e^{\frac{4\pi t}{M_S} \sigma}=e^{\frac{\mathcal{G}^0t}{M_S}}\left[\cos\left(\frac{4\pi  t}{M_S}\right)\mathbb{I}+\sin\left(\frac{4\pi  t}{M_S}\right)\sigma\right].
\end{equation}
As such, the inverse of the integration factor at large \(t\) behaves as
\begin{equation}
    \left(e^{\int^t\mathcal{P}_\infty~d\tilde{t}}\right)^{-1}=e^{-\frac{\mathcal{G}^0t}{M_S}}\left[\cos\left(\frac{4\pi  t}{M_S}\right)\mathbb{I}-\sin\left(\frac{4\pi  t}{M_S}\right)\sigma\right]
\end{equation}
which goes to zero when \(t\rightarrow\infty\). This factor suppresses the term with \(\dot{\textbf{R}}(0)\), which again, proves that the system of equation represents a limit cycle in momentum space. As mentioned in the main manuscript, the relaxation time is proportional to \(M_S\).
\subsection{B2: Properties of the limit cycle}
Let us denote the rotation matrix with rotation angle \(\zeta\) as
\begin{equation}
    \mathcal{R}(\zeta)=\begin{bmatrix}
        \cos\zeta&\sin\zeta\\-\sin\zeta&\cos\zeta
    \end{bmatrix}.
\end{equation}
Notice that the driving term can be decomposed into a rotation matrix operating on a constant vector,
\begin{equation}
    \textbf{p}(t)=\frac{\mathcal{R}(\varphi_0-\omega t)}{M_S}\begin{bmatrix}
        D^--4\pi J^+_Q\\D^++4\pi J^-_Q
    \end{bmatrix} ,
\end{equation}
and the large \(t\) limit of integration factor is actually a rotation matrix with exponential factor
\begin{equation}
     e^{\int^t\mathcal{P}_\infty~d\tilde{t}}=e^{\frac{\mathcal{G}^0t}{M_S}}\mathcal{R}\left(\frac{4\pi t}{M_S}\right).
\end{equation}
Using this relations, we can approximate the solutions of \(\dot{\textbf{R}}\) at large \(t\) as compositions of rotation
\begin{equation}
    \dot{\textbf{R}}\approx\frac{1}{M_S}\left[ e^{-\frac{\mathcal{G}^0}{M_S}t}\mathcal{R}\left(-\frac{4\pi t}{M_S}\right)\int_0^te^{\frac{\mathcal{G}^0}{M_S}t'}\mathcal{R}\left(\varphi_0+\left(\frac{4\pi }{M_S}-\omega\right)t'\right)dt'\right]\begin{bmatrix}
        D^--4\pi J^+_Q\\D^++4\pi J^-_Q
    \end{bmatrix}~.
\end{equation}
The integral above can be computed in a straightforward manner, giving us the following approximation (after taking only the leading terms in \(t\))
\begin{eqnarray}
    &&e^{-\frac{\mathcal{G}^0}{M_S}t}\mathcal{R}\left(-\frac{4\pi t}{M_S}\right)\int_0^te^{\frac{\mathcal{G}^0}{M_S}t'}\mathcal{R}\left(\varphi_0+\left(\frac{4\pi }{M_S}-\omega\right)t'\right)dt'\nonumber\\&\approx& M_S e^{-\frac{\mathcal{G}^0}{M_S}t}\mathcal{R}\left(-\frac{4\pi t}{M_S}\right)e^{\frac{\mathcal{G}^0}{M_S}t}\mathcal{R}\left(\varphi_0+\left(\frac{4\pi }{M_S}-\omega\right)t-\tan^{-1}\left(\frac{4\pi-M_S\omega}{\mathcal{G}^0}\right)\right).
\end{eqnarray}
The resulting large \(t\) approximation of the velocity is then given by
\begin{equation}
    \dot{\textbf{R}}\approx\mathcal{R}\left(\varphi_0-\omega t-\tan^{-1}\left(\frac{4\pi-M_S\omega}{\mathcal{G}^0}\right)\right)\begin{bmatrix}
        D^--4\pi J^+_Q\\D^++4\pi J^-_Q
    \end{bmatrix}~.
\end{equation}
We can observe that since the above factor is a rotation matrix that oscillates with frequency \(\omega\), the steady-state solutions of \(\dot{\textbf{R}}\) trace a circle with a certain radius, which we denote as \(v\). This radius can be numerically computed via the time-average \(\langle\dot{X}^2+\dot{Y}^2\rangle_t\) and taking \(t\) to be large such that the contribution from the transient state is suppressed. The confinement area of the Skyrmion in real space can also be estimated from \(v\). Since \(\textbf{R}\) is the integral of \(\dot{\textbf{R}}\), then the radius of the confinement area is approximately \(v/\omega\). In contrast to the confinement area which is determined by the steady state behaviour of the Skyrmion trajectory, the displacement is determined by its transient state. This displacement can be computed numerically using \(\tilde{\textbf{d}}\equiv\left(\langle X\rangle_t-X(0),\langle Y\rangle_t-Y(0)\right)\) such that the oscillating effect of the steady state is suppressed by the average which leaves us with only the effective displacement. To summarize, the transient state is characterized by the displacement \(\tilde{\textbf{d}}\) and the steady state is characterized by the average speed \(v\). These two parameters characterizes the resulting limit cycle of the Skyrmion trajectory. 



\end{document}